\documentclass[a4paper,fleqn,usenatbib]{mnras}

\usepackage{newtxtext} 
\usepackage{mathptmx}
\usepackage{float}

\usepackage[T1]{fontenc}
\usepackage{ae,aecompl}

\usepackage{pdflscape}

\usepackage{graphicx}	
\usepackage{amsmath}	
\usepackage{amssymb}	

\title[Metal-rich halo]{The metal-rich halo tail extended in $|z|$: a characterization with Gaia DR2 and APOGEE}

\author[E. Fern\'andez-Alvar et al.]{
Emma Fern\'andez-Alvar$^{1,2}$\thanks{E-mail: emmafalvar@gmail.com},
Jos\'e G. Fern\'andez-Trincado$^{3,4}$\thanks{E-mail: jfernandez@obs-besancon.fr},
Edmundo Moreno$^{2}$,
\newauthor
William J. Schuster$^{5}$,
Leticia Carigi$^{2}$,
Alejandra Recio-Blanco$^{1}$,
Timothy C. Beers$^{6}$,
\newauthor
Cristina Chiappini$^{7,9}$,
Friedrich Anders$^{7,9}$,
Bas\'ilio X. Santiago$^{8,9}$,
Anna B. A. Queiroz$^{8,9}$,
\newauthor
Angeles P\'erez-Villegas$^{10}$,
Olga Zamora$^{11}$,
D. A. Garc\'ia-Hern\'andez$^{11}$
\newauthor
and
Mario Ortigoza-Urdaneta$^{3}$
\\
$^{1}$Universit\'e Cote d'Azur, Observatoire de la Cote d'Azur, CNRS, 06300, Laboratoire Lagrange\\
$^{2}$Instituto de Astronom\'ia, Universidad Nacional Aut\'onoma de M\'exico, AP 70-264, 04510, Ciudad de M\'exico, M\'exico\\
$^{3}$Instituto de Astronom\'ia y Ciencias Planetarias, Universidad de Atacama, Copayapu 485, Copiap\'o, Chile\\	 
$^{4}$Institut Utinam, CNRS UMR6213, Univ. Bourgogne Franche-Comt\'e, OSU THETA , Observatoire de Besan\c{c}on, \\BP 1615, 25010 Besan\c{c}on Cedex, France\\ 
$^{5}$Instituto de Astronom\'ia, Universidad Nacional Aut\'onoma de M\'exico, AP 106, 22800 Ensenada, B. C., M\'exico\\  
$^{6}$Department of Physics and JINA Center for the 
Evolution of the Element (JINA-CEE),
University of Notre Dame,\\
 Notre Dame, IN 46556 USA\\
$^{7}$Leibniz-Institut fur Astrophysik Potsdam (AIP), An der Sternwarte 16, 14482 Potsdam, Germany\\
$^{8}$Instituto de F\'isica, Universidade Federal do Rio Grande do Sul, Caixa Postal 15051, 91501-970 Porto Alegre, Brazil\\
$^{9}$Laborat\'orio Interinstitucional de e-Astronomia--LIneA, Rua Gal. Jos\'e Cristino 77, 20921-400 Rio de Janeiro, Brazil\\
$^{10}$Universidade de S\~ao Paulo, IAG, Rua do Mat\~ao 1226, Cidade Universit\'aria, 05508-900, S\~ao Paulo, Brazil\\
$^{11}$Instituto de Astrof\'isica de Canarias, 38205 La Laguna, Tenerife, Spain
}

\date{Accepted XXX. Received YYY; in original form ZZZ}

\pubyear{}

\begin{document}
\label{firstpage}
\pagerange{\pageref{firstpage}--\pageref{lastpage}}
\maketitle

\begin{abstract}

We report an analysis of the metal-rich tail ([Fe/H] $> -0.75$) of
stars located at distances from the Galactic plane up to $|z| \sim
10$ kpc, observed by the Apache Point Observatory Galactic Evolution
Experiment (APOGEE). We examine the chemistry, kinematics, and dynamics
of this metal-rich sample using chemical abundances and radial
velocities provided by the fourteenth APOGEE data release (DR14) and
proper motions from the second Gaia data release (DR2). The analysis
reveals three chemically different stellar populations in the [Mg/Fe]
vs. [Fe/H] space -- a high-[Mg/Fe] and low-[Mg/Fe] populations, and a third group with intermediate [Mg/Fe]
$\sim$+0.1 -- as well as for other chemical elements. We find that they are also kinematically and
dynamically distinct. The high-[Mg/Fe] population exhibits a prograde
rotation which decreases down to 0 as $|z_{max}|$ increases, as well as eccentric orbits that are more bound and closer to the plane. The low-[Mg/Fe] stars are likely Sagittarius members, moving in less-bound orbits reaching larger distances from the centre and
the Galactic plane. The intermediate-Mg stars resembles the two stellar overdensities lying about $|z| \sim 5$ kpc
recently reported in the literature, for which a disc origin has been
claimed. We report the identification of new members of these two disc-heated overdensities.

\end{abstract}

\begin{keywords}
Galaxy -- halo -- stellar populations
\end{keywords}



\section{Introduction}

The bulk of the stars in the Galactic halo are characterized by low
metallicities, indicating that they were born during the first stages of
Galactic formation, from gas enriched by the first generations of stars.

The seminal monolythic collapse model of Eggen et al. (1962) predicted
that halo stars are the first stars born during the collapse of the
protogalactic cloud, before the gas settled and formed a disc. However,
within the current cosmological paradigm, the $\Lambda$ Cold Dark Matter
model describes galaxy formation as resulting from the assembly of larger
structures from mergers of smaller substructures. This implies that the
Galactic halo was populated by stars born in satellite galaxies which
underwent different, individual star-formation histories, and have been
accreted into the Milky Way over time. Observational evidence gathered
during the last decades provide support to this model, with the
detection of stellar density substructures accross the Galactic halo
(e.g., Grillmair 2009, Newberg \& Carlin 2015, Bernard et al. 2016).  

Recent studies of the Galactic stellar halo have provided signatures
for a variation in the spatial, chemical, and kinematical properties of
halo stars with distance from the Galactic Centre -- a break in the
spatial distribution (Deason, Belokurov \& Evans 2011; Ablimit \& Zhao
2018); a gradient in the mean iron abundance (which is usually
considered as a metallicity indicator in stars) with distance
(Fern\'andez-Alvar et al. 2015) correlated with the mean rotational
velocity component (Carollo et al. 2007, 2010; Beers et al. 2012; see
also the discussions in Sch\"onrich et al. 2014 and An et al. 2015); and
signatures of different chemical enrichment with distance
(Fern\'andez-Alvar et al. 2015, 2017; Lee et al. 2017).

Finally, two chemically distinct stellar populations in terms of the
$\alpha$-to-iron enrichment (Nissen \& Schuster 2010, 2011; Hawkins et al. 2015; Bergemann et al. 2017; Hayes et al. 2018a;
Fern\'andez-Alvar et al. 2018) were also detected.
All of these observational results led to the claim for a possible dual
formation scenario for the Galactic halo, comprising in-situ star
formation as well as accretion from satellite galaxies. Several numerical galaxy formation studies have considered the relative
contribution of in-situ star formation with respect to an accreted origin
throughout the halo (e.g., Font et al. 2011; Tissera et al. 2012, 2013,
2014), and found that this relative fraction affects the global chemical
and kinematical properties of the stars and their variation with
distance.

Although the majority of the halo stars have [Fe/H] $< -1.0$, some of
the previously mentioned works revealed stars at higher metallicities
that strongly differ in kinematics with respect to disc stars, and
instead exhibit halo-like motions (Nissen \& Schuster et al. 2010, 2011;
Schuster et al. 2012; Fern\'andez-Alvar et al. 2018). It is in this
metal-rich range, $-1.6 <$ [Fe/H] $< -0.4$, where the two distinct
[$\alpha$/Fe] populations were detected in the first place. 

Based on the kinematical and dynamical properties of these metal-rich
halo stars, and their overlap with the thick-disc chemical patterns in
the [$\alpha$/Fe] vs. [Fe/H] space, Nissen $\&$ Schuster (2010, 2011)
and Schuster et al. (2012) proposed that the high-$\alpha$ stars could
be old disc or bulge stars heated to halo kinematics by merging
satellite galaxies, or could have formed during the collapse of a
proto-Galactic gas cloud. Conversely, from their kinematics, the
low-$\alpha$ populations would have been accreted, presumably from dwarf
galaxies with low star-formation rates.

Bonaca et al. (2017) kinematically identified halo stars within 3 kpc
of the Sun moving with relative speeds larger than 220 km$ s^{-1}$ with
respect to the Local Standard of Rest (LSR). Surprisingly, half of this
kinematically-defined halo sample have [Fe/H] $> -1.0$. Because these stars move on orbits
that are preferentially aligned with the disc rotation, they proposed
that these metal-rich halo stars may have formed in-situ in the disc,
and later were perturbed by interactions with a more massive dwarf
galaxy, rather than being accreted from satellite galaxies.
These authors made use of the Gaia DR1 (Gaia Collaboration et al. 2016) and the Tycho-Gaia
solution (Michalik, D., Lindegren, L., \& Hobbs 2015) to derive parallaxes and proper motions, which lacked the
greater accuracy achieved in Gaia DR2.

Taking advantage of the enourmous improvement in accuracy and precision
provided by the Gaia DR2 data (Gaia Collaboration et al. 2018), Helmi et
al (2018) claimed an accretion origin for stars with halo-like motions
in the Solar Neighbourhood, which they associated with a relic of what
could be the last major merger event to have occurred in the Galaxy,
``Gaia-Enceladus''. Previously, Belokurov et al. (2018) had
detected a radial anisotropy in halo stars at [Fe/H] $> -1.7$, suggesting
that most of the inner, metal-rich halo would be dominated by stars
accreted from a single massive merger event. This idea was later
corroborated by several works (Myeong et al. 2018a,b, Deason et al.
2018, Lancaster et al. 2018, Simion et al. 2019).

Haywood et al. (2018) also supported this idea, from their analysis
of two overdensities in the Hertzsprung-Russell diagram (HRD)
revealed by Gaia Collaboration et al. (2018b), atributted to the halo
and the disc. They compared the two Nissen \& Schuster populations, also
using results from Gaia DR2, and found that these two groups of stars are
located over the two different sequences in the HRD, providing evidence
that they are in fact two different stellar populations (the situation
remained unclear for stars with [Fe/H] $< -1.0$). They also analysed the
kinematical and dynamical properties of the two groups
considering stars from the Apache Point Observatory Galactic Evolution Experiment
fourteenth data release (APOGEE DR14; Majewski et al. 2017, Abolfathi et
al. 2018), confirming differences between them, and argued for a likely disc origin for the
high-$\alpha$, metal-rich stars, and an accreted origin of the
low-$\alpha$ group and metal-poor stars, commensurate with the Helmi et
al (2018) results.

Recently, Mackereth et al. (2018) provided yet more evidence
from a detailed examination of the metal-poor stars in the APOGEE DR14
database. They found that 2/3 of nearby halo stars are characterized by
orbital eccentricities larger than 0.8, and chemical trends similar to
current massive dwarf galaxy satellites of the Milky Way. This group of
stars would presumably belong to the same accretion event. They also
exhibit slightly retrograde motions, and reach larger heights from the
plane and apocentric distances during their orbits. Interestingly, they
presented the same radial anisotropy as found by Belokurov et al. (2018)
. A comparison with the EAGLE suite of cosmological simulations
constrained the mass of such an accreted satellite to the range
$10^{8.5} \lessapprox M_{\ast} \lessapprox 10^{9} M_{\odot}$.

In this work we aim to analyse the metal-rich halo component at
distances beyond the Solar Neighbourhood. For this purpose we consider
the chemical, kinematical, and dynamical properties of a halo sample
selected from APOGEE DR14 combined with Gaia DR2 (Gaia Collaboration et al. 2016, 2018b). 
APOGEE is one of the Sloan Digital Sky Survey IV programs (Blanton et
al. 2017) devoted to exploration of the various Galactic components, in
order to understand the formation and evolution of our Galaxy. 

This paper is organized as follows. The APOGEE data and our criteria to
select halo stars are briefly explained in Section \ref{data}. In
Section \ref{analysis} we describe the chemical, kinematical, and
dynamical analyses performed and the results obtained. Section
\ref{discussion} is a discussion of the implications of our
results. Finally, we summarize our main conclusions in Section
\ref{conclusions}.

\section{Data}
\label{data}

We select our sample of halo stars from the APOGEE fourteenth data
release (DR14; Abolfathi et al. 2018). This data release includes data
from the SDSS-III (Eisenstein et al. 2011) APOGEE-1 program (which
comprises observations gathered from August 2011 to July 2014 -- see
Zasowski et al. 2013), as well as from the first two years of SDSS-IV
APOGEE-2 data (July 2014 - July 2016 -- see Zasowski et all. 2017).
Observations were taken using the Sloan Foundation 2.5-meter telescope
(Gunn et al. 2006) at Apache Point observatory, and the APOGEE
spectrograph (Wilson et al. 2010). Data for both programs was processed
(Nidever et al. 2015), and chemical abundances and radial velocities
derived (Holtzman et al. 2015), with the APOGEE Stellar Spectra
Parameter Pipeline (ASPCAP; Garc\'ia-P\'erez et al. 2016).

We choose field stars located at distances from the Galactic plane
larger than 5 kpc. This selection cut aims to favor the selection of halo against thick-disc stars. The scale height of the thick disc has been
estimated to be $\lessapprox 1$ kpc at the solar radius (see figure 8 in
Mateu \& Vivas 2018). The outer disc is warped and flared, and the scale
height could vary by a factor of two between the Solar Neighbourhood and
a distance of 12 kpc along the plane (Am\^ores et al. 2017). Mateu \&
Vivas (2018), by comparing with a sample of RR-Lyraes, estimated no
thick-disc stars being found at $|z| >$ 3 kpc at the solar radius. Using the Galactic
potential model 
\texttt{GravPot16}
(Fern\'andez-Trincado et al. 2019, in prep.) that we describe in Section 2.2
below, we estimate that the number of particles at |z| > 5 kpc with respect to the
total of particles at more than 4 kpc from the Galactic Center is 2.5\%
for the thick disc and 47.4\% for the halo. 

However, Juric et al. (2008), using low-resolution SDSS data, and
more recently, Wang et al (2018) from low-resolution LAMOST
data, fit models showing that the stellar densities of halo and
thick-disc stars at $|z| \sim 5$ kpc are still comparable. Juric et al.
(2008) classified thick-disc and halo stars by their colours as a proxy
for metallicity, considering halo stars those with [M/H] $< -1.0$. Under the light of these works 
it is not clear how negligible is the contribution of stars moving in thick-disc like orbits at such distances.

The above studies also show that the  disc-halo transition is still not fully understood. 
In this framework, our target selection cut will help to better
characterize the chemical, kinematical, and dynamical properties of stars
at |z| > 5 kpc, and help to unveil their origins.

We reject stars with flags warning of poor stellar parameter estimates,
and those with signal-to-noise ratios lower than 80. We also only
consider stars with 1.0 $< \log g < 3.5$ and $T_{\rm eff} > 4000 $\,K,
for which the derived chemical abundances are more accurate (see
Holtzmann et al. 2018, J\"onsson et al. 2018, Zasowski et al. 2018). In
the following analysis we calculate orbital parameters using Gaia DR2
absolute proper motions ($\mu_{\alpha}$ and $\mu_{\delta}$) and radial velocities
($v_{rad}$) released by APOGEE DR14. The median and
MAD\footnote{Median Absolute Deviation} of the relative errors in proper
motions are $2\pm1\%$ and $1\pm1\%$ for $\mu_{\alpha}$ and
$\mu_{\delta}$, respectively, and $0.2\pm0.2\%$ for $v_{rad}$. Our
final sample comprises a total of 504 stars. The coordinates,
stellar parameters, [Fe/H] and [Mg/Fe] abundances, as well as the radial
velocities, parallaxes, proper motions, and distance estimates for this
sample are shown in Tables \ref{tab:params1} and \ref{tab:params2}.
This sample covers a distance to the Galactic Centre, $|d_{GC}|$,
between 5 and $\sim$ 25 kpc. Several works have pointed out that $\sim
15-20$ kpc is a transition radius between the inner- and outer
halo-regions (Carollo et al. 2007, 2010; Tissera et al. 2014;
Fern\'andez-Alvar et al. 2017). Thus, with our sample, we examine
primarily the inner-halo region.

\subsection{Distances}

\begin{figure}
	\includegraphics[scale=0.55, trim= 0 0 0 20]{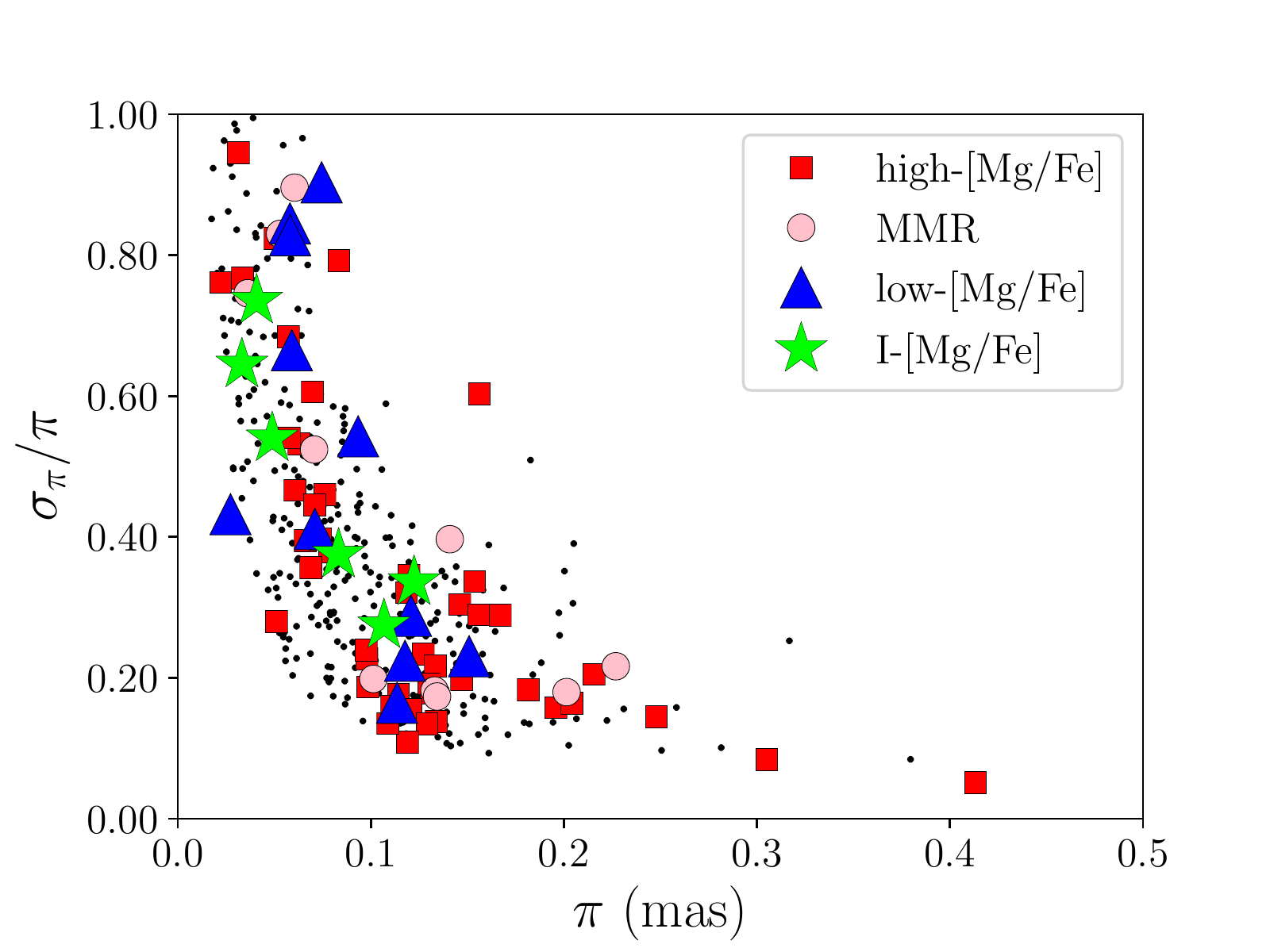}
	\includegraphics[scale=0.55, trim= 0 0 0 0]{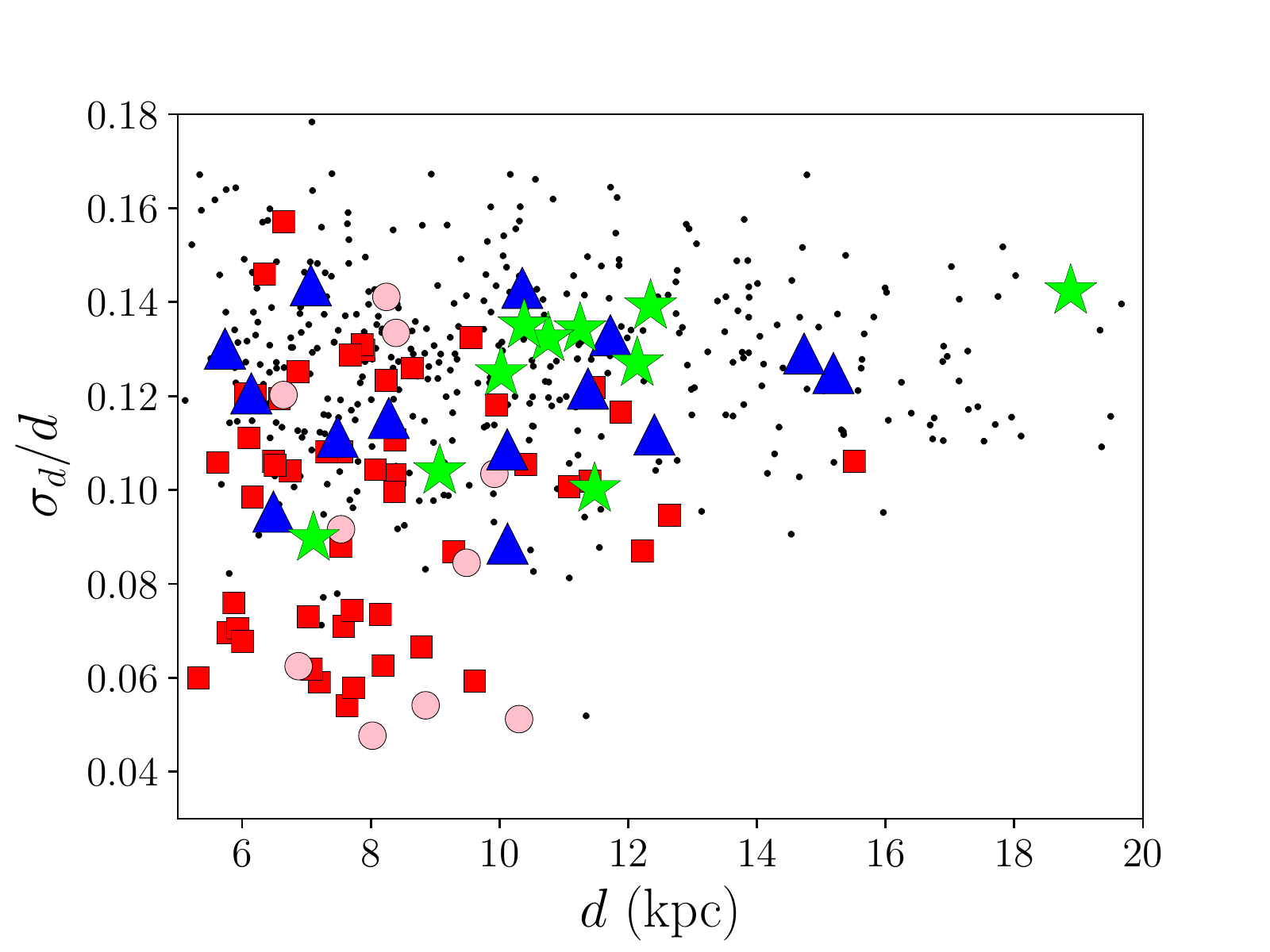}
        \caption{Top panel: Relative errors in Gaia DR2 parallaxes. Bottom panel: relative errors in the BPG distance estimates.}
    \label{fig:dist_unc}
\end{figure}

Distances were calculated by the SDSS Brazilian Participation Group
(BPG) using the StarHorse code described in Santiago et al. (2016) and
Queiroz et al. (2018). The code calculates the posterior probability
distribution over a grid of stellar evolutionary models (PARSEC 1.2S --
Bressan et al. 2012) using a Bayesian methodology similar to that of
Binney et al. (2014). The input parameters are the spectroscopic
parameters measured by ASPCAP, Gaia DR2 parallaxes corrected for a
global zero-point shift of $-0.052$ mas (as determined by Zinn et al.
2018), as well as APASS DR9 $BVgri$, 2MASS $JHK_s$ and WISE W1W2
photometry. Thus, the distance is determined by marginalising
the joint posterior PDF, not only from the parallax information, but
from spectroscopic parameters and magnitudes as well. For more details,
see sections 2 and 3 in Queiroz et al. (2018). The combination of Gaia
DR2 and APOGEE data within the StarHorse code is described in detail in
Santiago et al. (in prep.).

The priors include the space density, IMF, age distributions, and MDF
of the main Galactic components (thin and thick disc, bulge, and halo).
The adopted priors are as in Binney et al. (2014) (with updates in some
structural parameters, from Bland-Hawthorn \& Gerhard 2016 -- see
Queiroz et al. 2018, section 3.3), and additional spatial priors for the
bulge/bar. The age and metallicity priors are Gaussians for each
Galactic component. As explained in Queiroz et al., this choice was
made for simplicity, and in order to include most of the recent age and
MDF found in the literature, but avoiding making the priors too
especific. They assume that the impact of this choice is minor. For more
details see section 3.3 in Queiroz et al. 2018.

The top panel in Figure \ref{fig:dist_unc} shows the fractional parallax 
uncertainties, and the bottom panel displays the relative errors of the
distance estimates (the standard deviation divided by the mean of the
probability distribution function) for our selected sample (we display
stars with [Fe/H] $> -0.75$ with  different symbols, because in the
following analysis we classify stars in different groups, as we
explain in Section 3.1.2). Although the relative errors in parallax
are very large, the errors in distances are estimated to be less than
the $20\%$. 

We quantify whether this is likely because in these cases the spectroscopic parameters and magnitudes primarily influence the distance determination. We compare them with the recent spectro-photometric distances computed by Leung \& Bovy 2019, which do not consider Gaia parallaxes as a prior. We see that, considering the whole sample, 60\% of the stars show differences of less than the 30\%. For metal-rich stars ([Fe/H] $> -0.75$, the group that we will focuse on in this work) 80\% of them show differences of less than the 20\%. The few stars with larger differences are not dominant in any of the different chemical groups analysed later.

We compared the BPG distances with those calculated by Bailer-Jones et al. (2018)
and McMillan (2018), which also used Gaia DR2 parallaxes. In both
cases, the calculated distances are from a Bayesian procedure, but the first one
considers a single purely geometrical prior, while the second one also
includes the radial velocity as a prior. We found a large dispersion in
the comparison with the Bayler-Jones distances, although most of our
stars show relative differences lower than the $40\%$. There are only 81
stars with McMillan distance estimates, but the comparison is better,
showing relative differences less than $20\%$. 

We repeated the analysis explained in the following sections using three different
samples: (\textit{i}) our full stellar selection, (\textit{ii}) stars
with relative differences with the Bayler-Jones estimates less than
$40\%$, and (\textit{iii}) stars with fractional parallax uncertainties
less than $50\%$. Tables \ref{tab:kinematics}, \ref{tab:dynamics1}, and
\ref{tab:dynamics2} show the median and median absolute deviation
obtained for velocities and orbital parameters. From inspection,
the values are similar.  We check that the distributions of the
velocities and orbital parameters do not substantially change and, consequently,
neither do the conclusions of this paper. For this reason, we maintain the
analysis performed over the full sample.

\subsection{The Galactic models}

To provide a comprehensive study of our sample, we perform an orbital analysis. In order to do that, we use two non-axisymmetric models 
for the Galactic gravitational potential. These models are realistic (as far as possible) potentials, that fits the dynamical parameters to our best current knowledge of the Milky Way.

The first non-axisymmetric model, \texttt{GravPot16}\footnote{\url{https://gravpot.utinam.cnrs.fr}}
(Fern\'andez-Trincado et al. 2019, in prep.), considers a 3D steady-state
gravitational potential for the Galaxy, modelled as the sum of
axisymmetric and non-axysimmetric components. The axisymmetric component
is made-up of the superposition of seven thin discs with Einasto laws
(Einasto 1979), two thick discs with sech$^{2}$ laws (Robin et al.
2014), the ISM component with a density mass and a Hernquist stellar halo mass distribution like
that in Robin et al. (2014). The non-axisymmetric component is a 
boxy/peanut bar with an assumed mass, present-day orientation and pattern speeds, which is within observational estimates that lie in the range of 1.1$\times$10$^{10}$ M$_{\odot}$, 20$^{\circ}$, and 30--50 km s$^{-1}$ kpc, respectively. The Sun is placed at 8.3 kpc, the local 
rotation velocity is $-239$ km s$^{-1}$, and the Sun's orbital velocity vector $[U,V,W]_\odot$ = [$-11.1$,
$-12.24$, 7.25] km s$^{-1}$ (Sch\"onrich et al. 2010, Brunthaler et al. 2011). \texttt{GravPot16} has been adopted in a number of recent papers (see, e.g., Fern\'andez-Trincado et al. 2016b, 2017; Recio-Blanco et al. 2017; Albareti et al. 2017; Helmi et al. 2018, Schiappacasse-Ulloa et al. 2018; Tang et al. 2018, Contreras Ramos et al. 2018, Fern\'andez-Trincado et al. 2019a). For a more detailed discussion, we refer the reader to a forthcoming paper (Fern\'andez-Trincado et al. 2019, in prep.).

By combining \texttt{GravPot16} with precise information from radial velocity, absolute proper motion, distance, and sky position, we ran 10$^{5}$ orbit simulations for each of the metal-rich halo stars, taking into account the uncertainties in the input data, where the errors were propagated as $1\sigma$ variations in a Gaussian Monte Carlo re-sampling. For each generated set of parameters, the orbit is computed backward in time up to 2.5 Gyr. From the integrated set of orbits, we compute the median values for (\textit{i}) r$_{peri}$ and r$_{apo}$, the peri-/apo-galactocentric radius, respectively; (\textit{ii}) the orbital eccentricity, defined as \textit{e} = (r$_{apo}$ $-$r$_{peri}$)/(r$_{apo}$ + $r_{peri}$), (\textit{iii})  the total orbital energy $E$, (\textit{iv}) the z-component of the angular momentum in the inertial frame, $L_z$, and (\textit{v}) the maximum vertical excursion from the Galactic plane, $|z_{max}|$. The uncertainty ranges of the orbital elements are given by the 16$^{th}$ and 84$^{th}$ percentile value.

The second Galactic potential is based on the axisymmetric model of Allen \& Santillan (1991; AS91) using the Galactic prolate bar model (\textit{non-axisymmetric} contribution) as described in Pichardo et al. (2004). The total potential has been rescaled to the Sun's Galactocentric distance and LSR velocity employed in AS91. The present orientation of the major axis of the Galactic bar is taken as $20^{\circ}$, and the angular rotation speed of the bar is 45 km s$^{-1}$kpc$^{-1}$. The results obtained  with this model are consistent with \texttt{GravPot16}. Therefore, this second model was used to show the robustness of our analysis.

\section{Analysis}
\label{analysis}

\subsection{Chemistry}

\begin{figure*}
	\includegraphics[scale=0.7, trim= 0 0 0 20]{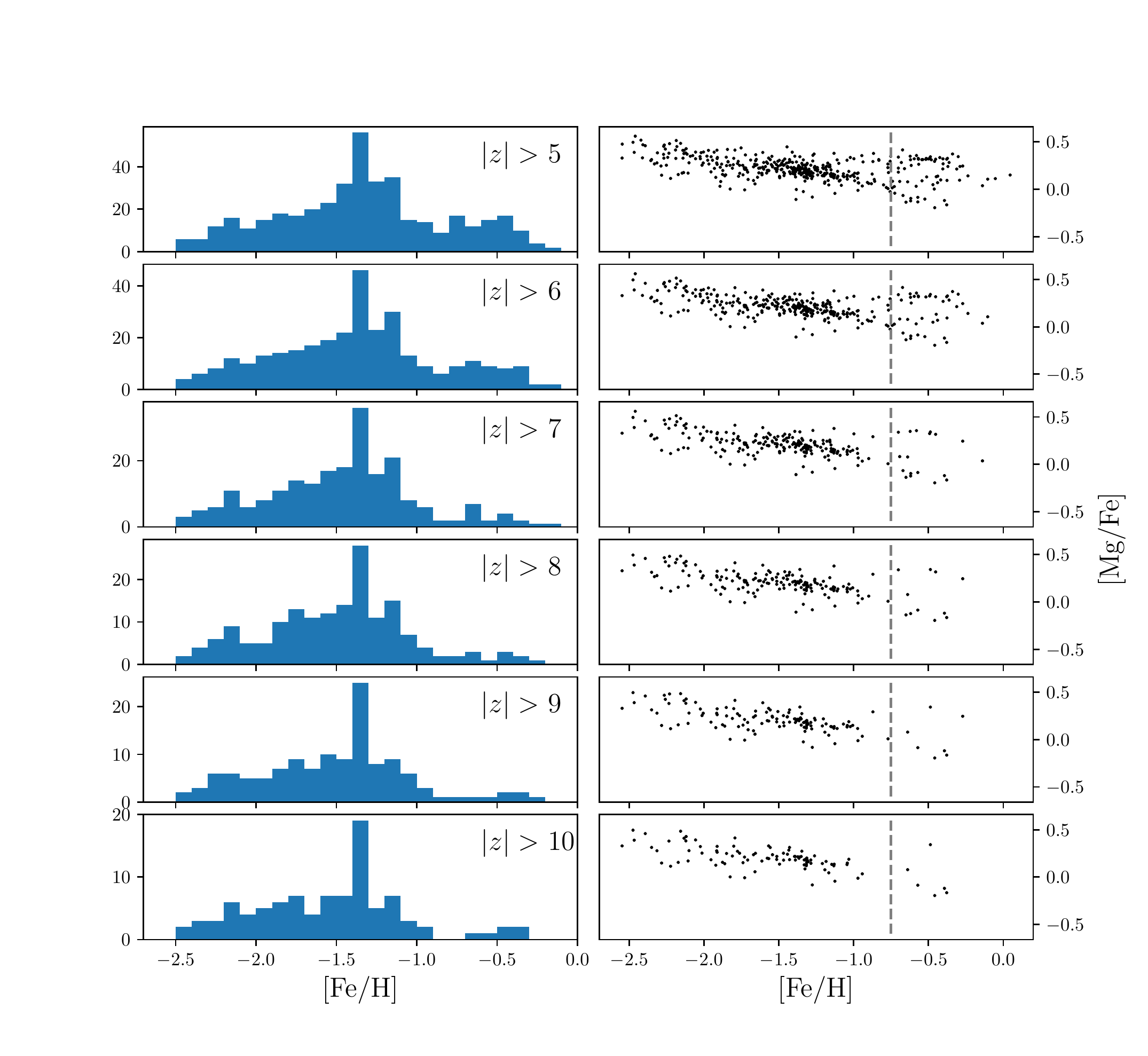}   
    \caption{Left panels: Metallicity distribution functions for several
    subsamples of our selected APOGEE halo stars, considering different
    $|z|$ cuts. Right panels: [Mg/Fe] vs. [Fe/H] display for the
    corresponding subsample on the left. A vertical dashed line divides our halo
sample into metal-poor and metal-rich subsamples at [Fe/H] $= -0.75$.} \label{fig:mdf_mgfe}
\end{figure*}

\subsubsection{The MDF as a function of z.}

We first examine the metallicity distribution function (MDF) as a function of
the distance from the Galactic plane, $|z|$. The  left panels in Figure
\ref{fig:mdf_mgfe} show the MDF for stars in our sample, but considering
different cuts in $|z|$, i.e., those at distances $|z| >$ 5, 6, ..., 10
kpc. The resulting distributions exhibit peaks at [Fe/H] $\sim -1.4$ and
extended tails at both lower and higher [Fe/H], reaching up to solar
values. In particular, there is a sample of stars at metallicities
greater than [Fe/H] = $-0.75$, with a peak at [Fe/H]$\sim -0.5$, which 
remains even at distances $|z| >$ 10 kpc (although at  |z| > 8 kpc the high-[Mg/Fe] population is very reduced, within the Poisson noise of the sample). 

We evaluate possible selection effects due to the different cuts in
color, magnitude, $T_{\rm eff}$ and $\log g$ performed to obtain our
halo sample. We reproduce these selection cuts (excluding also stars at
$|z| < 5$ kpc) over a TRILEGAL simulation (Girardi et al. 2012),
centered at $(l,b)=(60, 75)$. This analysis reveals that the restriction
to the 1.0 $< \log g <$ 3.5 range distorts the MDF, and a peak at high
metallicities, larger than $-$1.0, appears. For this reason we cannot rely
on the shape of the MDF on the metal-rich side that we obtain from
observations; however, this does not change the fact that we are
detecting metal-rich stars at high distances from the plane. 

The peak at [Fe/H] $= -1.4$ differs from the typical value associated
with the MDF of the inner halo (see Carollo et al. 2007; Allende Prieto
et al. 2014). We have checked if this could be due to the method
employed to determine chemical abundances by the ASPCAP pipeline. The
stars do not exhibit any particular trend in $T_{eff}$ and $\log g$, and
homogeneously cover the stellar parameter ranges for the entire halo
sample. Even if this clump of stars at [Fe/H] $= -1.4$ is related to
some systematic effect by the abundance determinations that we are not
able to detect, it does not affect our main results, since in our
following analysis we focus on stars with [Fe/H] $> -0.75$. 

The right panels of Figure \ref{fig:mdf_mgfe} show the stars in the
chemical space [Mg/Fe] vs. [Fe/H] for
each sub-sample considering the different $|z|$ cuts. Those stars at
[Fe/H]$ > -0.75$ are split into different sequences of [Mg/Fe] as a
function of [Fe/H].  A significant number of stars exhibit a constant trend
of larger [Mg/Fe] values, whereas another subsample of stars follow the
decreasing trend with [Fe/H] observed for the main halo sample at lower
metallicities, reaching sub-solar values.  Finally, there are stars
populating the space between both sequences, some of them reaching the
largest [Fe/H] values in the whole sample, [Fe/H] $> -0.3$.

\begin{figure*}
	\includegraphics[scale=0.6]{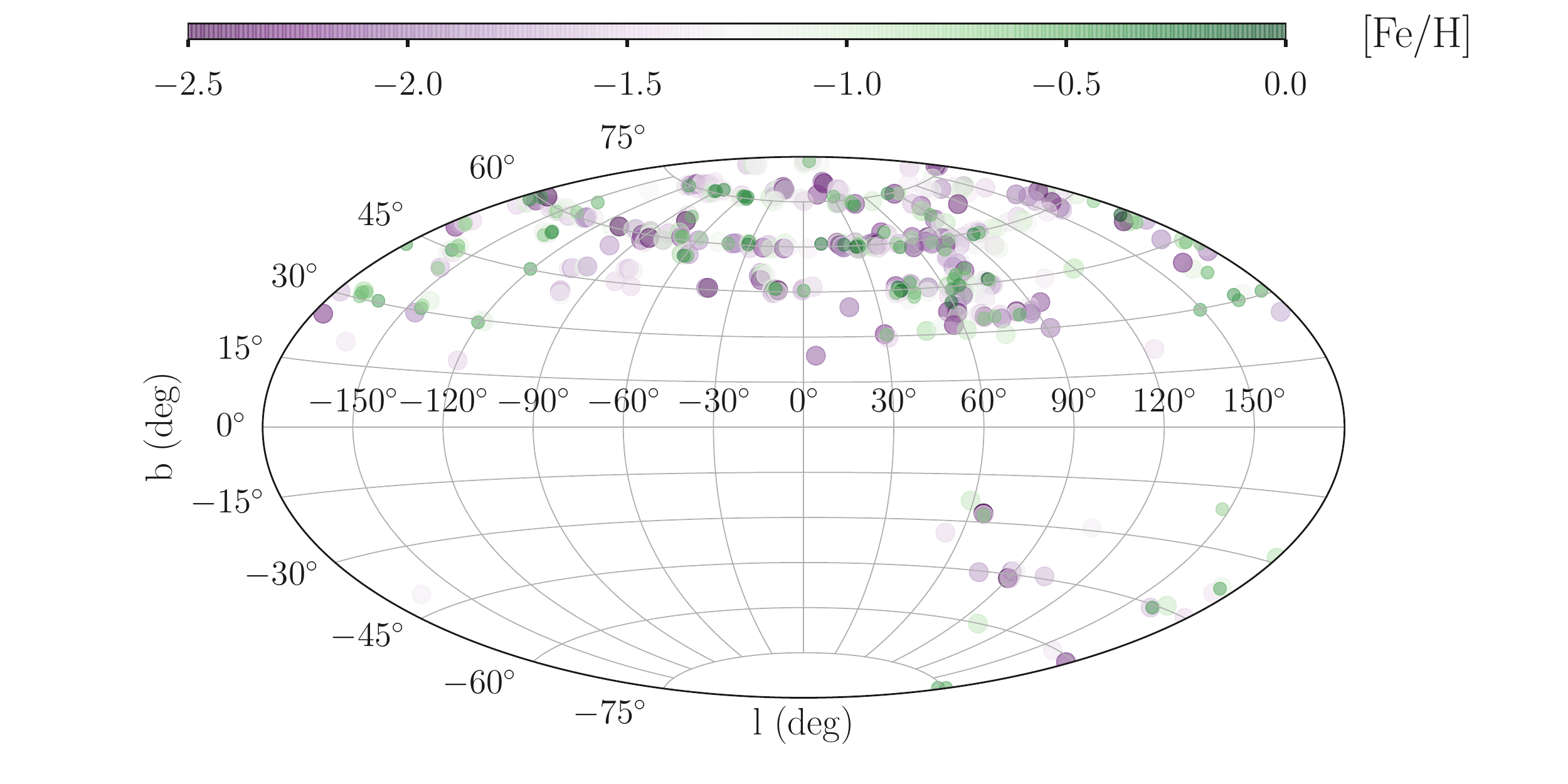}
        \caption{Distribution of our halo sample in Galactic coordinates on an Aitoff projection. The stars are colour-coded by [Fe/H].}

    \label{fig:aitoff}
\end{figure*}

\begin{figure}
	\includegraphics[scale=0.55]{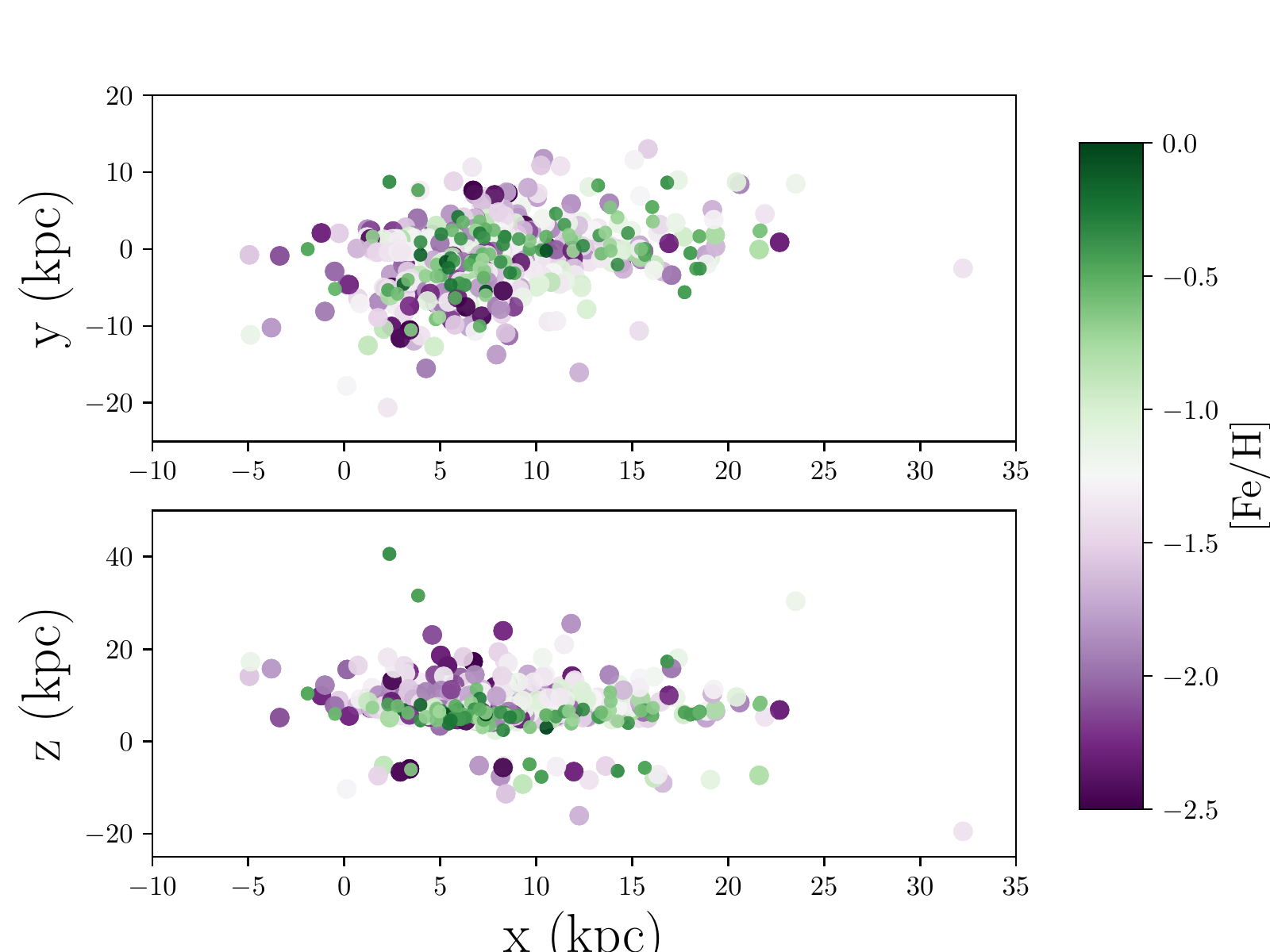}
        \caption{Distribution of our halo sample in
    Galactic $xy$ and $xz$ coordinates. ($x$,$y$,$z$) form a
    right-handed system, with $x$ positive towards the Sun, $z$ towards
    the North Galactic Pole, and $y$ in the direction opposite to Galactic
    rotation. The colour scale indicates the [Fe/H] of each star, as in
    Figure \ref{fig:aitoff}.} 
\label{fig:xyz}
\end{figure}

Figure \ref{fig:aitoff} shows the distribution of our sample in Galactic
coordinates on an Aitoff projection. Stars in the top and bottom
panels are depicted with colours indicating their [Fe/H]. Our sample, in
particular the metal-rich tail, is well-spread over the sky, and
exhibits no spatial evidence for being a recent accretion event. Figure \ref{fig:xyz} shows the stars in Galactic $xy$ (top panel)
and $xz$ (bottom panel) coordinates, with colours indicating their
[Fe/H]. Inspection of the panels shows that the metal-rich population
is located at low $|z|$, extending over a wide range of $x$. As $|z|$
increases the metal-poor halo begins to dominate. 

\subsubsection{The chemical characterization of the metal rich tail.}

For further analysis, we split our sample of stars at [Fe/H] $< -0.75$
(metal-poor) and at [Fe/H] $> -0.75$ (metal-rich). We classify the latter
considering their different [Mg/Fe] trends with [Fe/H]. We employed the
same division obtained by Hayes et al. (2018). They statistically
inferred, for a sample of APOGEE DR13 metal-poor stars, [M/H] $< -1.0$,
two [$\alpha$/Fe] vs. [Fe/H] sequences above and below a gap at [Mg/Fe]
$= -0.2x$[Fe/H]. Their sample comprises stars at lower [Fe/H] than ours,
but the work of Fern\'andez-Alvar et al. (2018) applied this
classification to a kinematically classified halo sample that covers
higher metallicities, [Fe/H] $\sim$ 0.0. The Hayes et al. division
cleanly separated stars into the two different [Mg/Fe] vs. [Fe/H] trends
that the sample displayed. 

Thus, we follow the same classification. However, there is a group of stars in our sample
located above the division line. A later analysis of the derived orbital
parameters revealed a group of intermediate [Mg/Fe] stars that largely
differs from the other stars. Consequently, we classify those stars as a
different group. In addition, at [Fe/H] $> -0.3$ stars show lower
[Mg/Fe] values; we analyse those stars separately. Thus, the final four
groups follow this classification: 

\begin{enumerate}

\item High-[Mg/Fe]: $-0.75 <$ [Fe/H] $< -0.3$ and [Mg/Fe]$ > +0.15$
\item Low-[Mg/Fe]: $-0.75 <$ [Fe/H] $< -0.3$ and [Mg/Fe] $< -0.3-0.65*$[Fe/H]
\item Intermediate-[Mg/Fe] (I-[Mg/Fe]): $-0.55 <$ [Fe/H] $< -0.3$
and $-0.05 <$ [Mg/Fe] $< +0.15$
\item the most metal-rich (MMR): [Fe/H]$ > -0.3$.
\end{enumerate}

In order to ensure that the differences in [Mg/Fe] observed for the different
stellar groups are physical, we investigate possible systematic errors
in the abundance determinations. We compare the [Mg/Fe] in each group
with the stellar parameters $T_{\rm eff}$, $\log g$ and microturbulence velocity.
These parameters also play a role in the formation of the spectral lines
and the under- or over-estimation of any of them can affect the derived
chemical abundances. Low-[Mg/Fe] stars show a correlation with $T_{\rm eff}$, $\log g$ and an anticorrelation with $v_{micro}$, suggesting that the lower abundance values could be underestimated in cool stars. To validate the [Mg/Fe] values,  we have compared the low-[Mg/Fe] spectra with those of I-[Mg/Fe] and high-[Mg/Fe] twins (i.e., diferences in their measured $T_{\rm eff}$, $\log g$, [Fe/H] and $v_{micro}$ lower than 80 K, 0.34 dex, 0.13 dex and 0.2 $km s^{-1}$, respectively). This comparison shows that the Mg lines used to derive the abundances are in fact weaker for the low-[Mg/Fe] stars, confirming their lower abundance. No other correlations were found between the [Mg/Fe] with any other parameter, indicating that the
chemical differences of the stellar groups are not due to systematic
effects.

\begin{figure*}
	\includegraphics[scale=0.75, trim= 0 10 0 15]{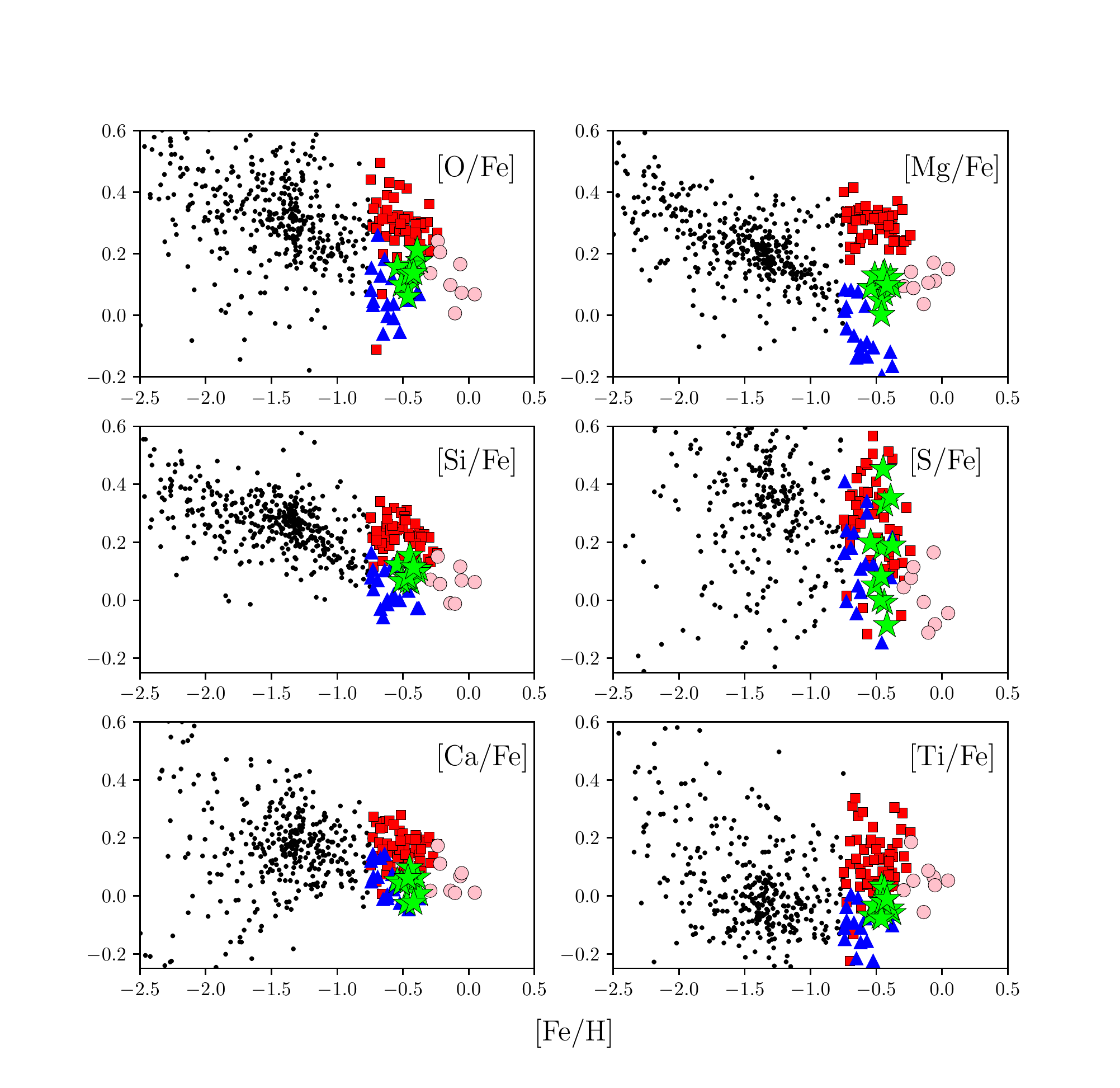}
    \caption{[X/Fe], as a function of [Fe/H], for the $\alpha$-elements
O, Mg, Si, S, Ca, and Ti. The metal-poor sample is plot with black dots, whereas the metal-rich stellar groups are depicted with the same symbols as in Figure \ref{fig:dist_unc}.}
   \label{fig:chem_alpha}
\end{figure*}

\begin{figure*}
	\includegraphics[scale=0.75, trim= 0 10 0 15]{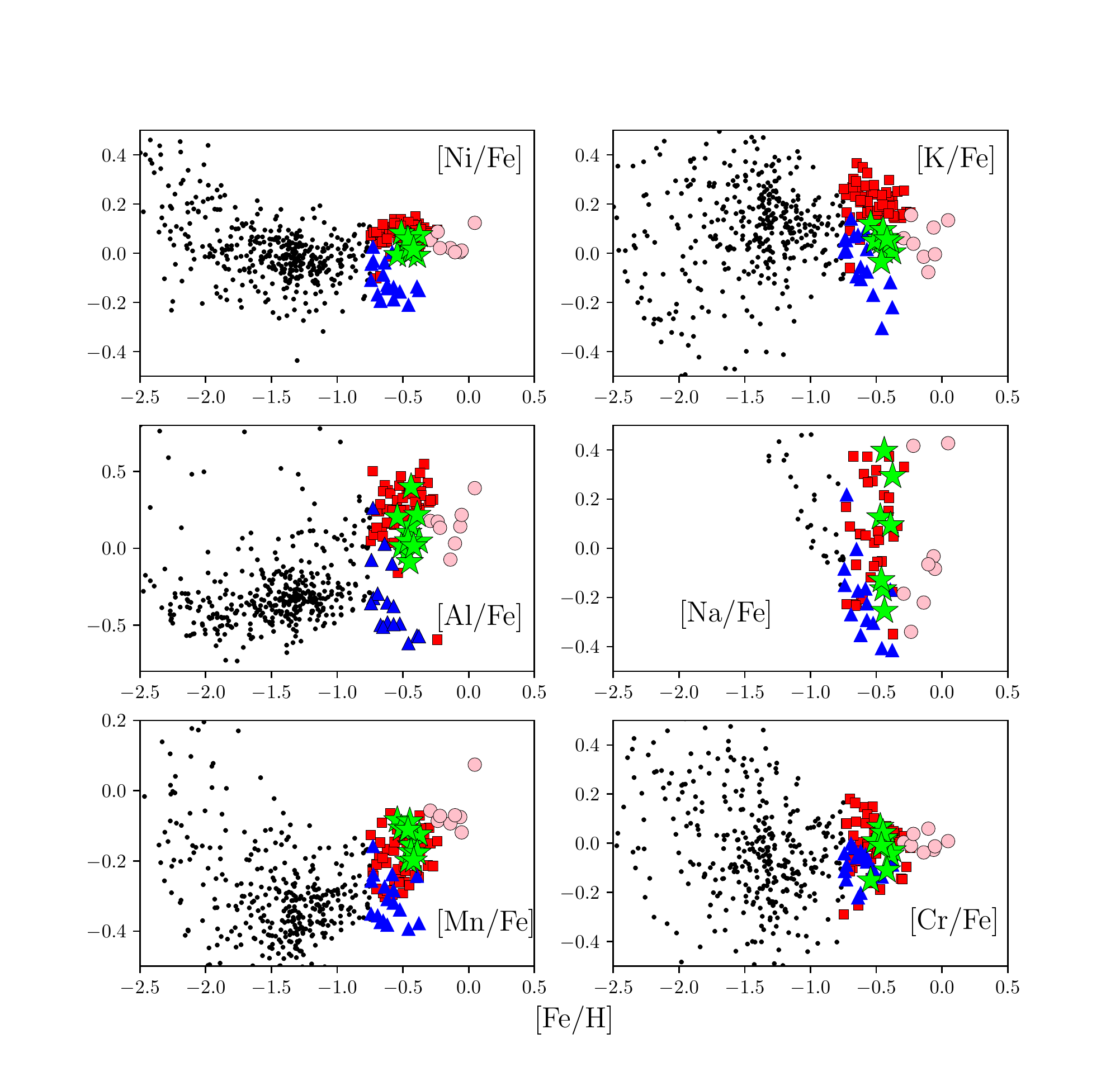}
    \caption{[X/Fe] as a function of [Fe/H] for Ni, K, Al, Na, Mn and Cr. The metal-poor sample is plot with black dots, whereas the metal-rich stellar groups are depicted with the same symbols as in Figure \ref{fig:dist_unc}.}
   \label{fig:chem_alpha}
\end{figure*}

We also explore other chemical species. The panels in Figure \ref{fig:chem_alpha} 
show the chemical trends of $\alpha$-element-to-iron ratios: [O/Fe],
[Mg/Fe], [Si/Fe], [S/Fe], [Ca/Fe], and [Ti/Fe] as a function of [Fe/H].
The largest differences in the chemical trends are observed in [Mg/Fe],
but for the other $\alpha$-elements the high-[Mg/Fe] (and low-[Mg/Fe])
stars also show higher (and lower) values than the I-[Mg/Fe] and MMR
stars. As observed by Nissen \& Schuster (2010), the differences are
milder for Ca than for the other $\alpha$-elements.

It is worth to emphasize that \texttt{ASPCAP} pipeline provides chemical abundance ratios under the assumption of local thermodynamic equilibrium (LTE), and therefore possible non-LTE (NLTE) and/or 3D effects on the derived values of the atmospheric parameters and chemical abundances are not taken into account. There are hints of these effects in some APOGEE stars (see Hawkins et al. 2016, for instance).

Bergemann et al. 2017b estimated that <3D> NLTE-corrections in metal-poor giants should be of the order of $\sim$0.1--0.2 dex, in the $T_{eff}$ regime of our objects from measurements in the optical range. However, they found that their resulting [Mg/Fe] distribution with [Fe/H] taking into account <3D> NLTE effects is consistent with the one obtained by ASPCAP from estimates in the infrarred H-band considering LTE, mainly at high metallicities (see Fig. 5d and Fig. 6 in Bergemann et al.). This encourages thinking that these effects in ASPCAP measurements for our stars should be small. In addition, there is a recent study by Zhang et al. (2017) which evaluated the NLTE effects in Mg lines in the H-band. The resulting [Mg/Fe] abundance differences considering LTE and NLTE for the thirteen stars that they compared were lower than 0.1 except for one star, and in the worse case the correction reaches up to $\sim-0.22$. The [Mg/Fe] values displayed by our high-[Mg/Fe] and low-[Mg/Fe] exceed this value, suggesting that the bimodality is not explained only by these effects. We are aware of that a <3D> NLTE analysis is important to well-establish the chemical differences between stellar populations. But this kind of analysis is beyond the scope of this paper.

Figure \ref{fig:chem_alpha} displays the trends with [Fe/H] for other elements
derived by ASPCAP, with a different nucleosynthetic origin: [Ni/Fe],
[K/Fe], [Al/Fe], [Na/Fe], [Mn/Fe], and [Cr/Fe]. As in previous works, the
dichotomy between high-[Mg/Fe] and low-[Mg/Fe] stars is also observed in
our data for [Ni/Fe], and also [Al/Fe] and [K/Fe] (see Nissen \&
Schuster 2010; Fern\'andez-Alvar et al. 2017). Nissen \& Schuster
detected the same dichotomy in [Na/Fe], however the ASPCAP [Na/Fe]
estimates exhibit a very large dispersion and no clear trend with
[Fe/H]. These estimates can be affected by strong telluric absorption
near the abundance windows in the stars (Zasowski et al. 2018). Thus,
they should be considered with caution. Nissen \& Schuster (2011) found
that their high- and low-$\alpha$ populations were indistinguishable in
[Mn/Fe] and [Cr/Fe] trends with [Fe/H]. Our results show that all the
populations overlap, although the low-[Mg/Fe] population tends to have
lower [Mn/Fe] and [Cr/Fe] values.

The I-[Mg/Fe] and MMR groups show values in between the high- and low-[Mg/Fe] 
stars for all the $\alpha$-elements (except [S/Fe]), and [Ni/Fe],
[K/Fe], and [Al/Fe]. In the case of [Mn/Fe] and [Cr/Fe], the I-[Mg/Fe] overlap
with the high-[Mg/Fe] group, and the MMR stars follow the same trend as the
high-[Mg/Fe] stars with [Fe/H]. 

The differences observed in the abundance of several elements, some of which previous studies also detected a dichotomy in their abundances,
underscore the conclusion that the stars that we are analysing at high
metallicities are chemically diverse.

\subsubsection{Belonging to known streams?}

In order to test if some of the metal-rich halo we identify belongs to a known
stream, we compare the spatial distribution of our halo sample with the
distances and positions of all the streams currently detected in the
halo. Some of them match the path of the Sagittarius stream in Galactic
coordinates, and some stars in our sample are located at distances
compatible to its trailing arm. At [Fe/H] $> -0.75$, Sagittarius stars
exhibit similar [Mg/Fe] values (Hasselquist et al. 2017) as most of our
low-[Mg/Fe] metal-rich halo stars. We also compared the abundance
ratios [O/Fe], [Si/Fe], [Ca/Fe], [Ni/Fe], [Al/Fe], [K/Fe], and [Mn/Fe],
and find that a number of stars (those with [Mg/Fe] $< 0.0$) in our low-[Mg/Fe] halo sample agree in those
abundance ratios with those measured for Sagittarius stars at the same
metallicities.

Comparison of the GRV \footnote{The radial velocity $v_{rad}$ corrected
for solar motion (we adopt the solar Galactocentric velocities
$U_{\odot}=11.1$ km$ s^{-1}$, $V_{\odot}=12.24$ km$ s^{-1}$ and $W_{\odot}=7.25$ km$ s^{-1}$,
as in Sch\"onrich et al. 2010).} (in the range [$-100$,100] km$ s^{-1}$) with
Carlin et al. (2018) shows an agreement with their GRV measurements for
Sgr. For these reasons, we conclude that stars with [Mg/Fe] $< 0.0$ in our low-[Mg/Fe] population are likely
associated with the Sagittarius stream.

We notice that the I-[Mg/Fe] group matches in ($l$,$b$), as well as in
chemistry, with the two stellar overdensities in Triangulum/Andromeda
(Rocha-Pinto et al. 2004; Martin et al. 2007; Price-Whelan et al. 2015;
Perottoni et al. 2018) and A13 (Sharma et al. 2010; Li et al. 2017).
These have been recently spectrocopically followed up by Bergemann et al.
(2018) and Hayes et al. (2018b). The latter analyzed APOGEE stars
specifically targeted to explore the Triaungulum/Andromeda overdensity.
However, none of our I-[Mg/Fe] stars are one of these targets. In the
following kinematical and dynamical analysis we will examine whether our
I-[Mg/Fe] stars belong to the same stellar population as these two
overdensities, and discuss the implications of our results for this
explanation of their origin.

In summary, our sample of halo candidates observed by APOGEE with $|z| > 5$ kpc
and 5 $< |d_{GC}| <$ 25 kpc ranges in metallicity over $-2.5 < $ [Fe/H]
$ < 0.0$. These stars exhibit a main decreasing trend of [Mg/Fe]
(and other $\alpha$-elements) with [Fe/H], from the most
metal-poor stars up to [Fe/H] $\sim -1.0$, where a split in several
trends is evident. In other words, the metallicitiy distribution
function of the stars populating the region at more than 5 kpc from the
Galactic plane shows a metal-rich tail, comprised of several
chemically distinct stellar populations, pointing to different
origins.

A dichotomy in the [$\alpha$/Fe] vs. [Fe/H] space had already been detected
in several previous works (Nissen $\&$ Schuster 2010; Recio-Blanco et al. 2014 ; Fern\'andez-Alvar
et al. 2018; Hayes et al. 2018a). However, this is the first time the
distribution of this metal-rich tail with $|z|$ is explored. Our sample
reveals that the split into chemically distinct stellar populations
reaches distances up to $|z| > 5$ kpc from the Galactic plane.

\subsection{Kinematics}

\begin{figure*}
	\includegraphics[scale=0.8, trim=100 0 100 0]{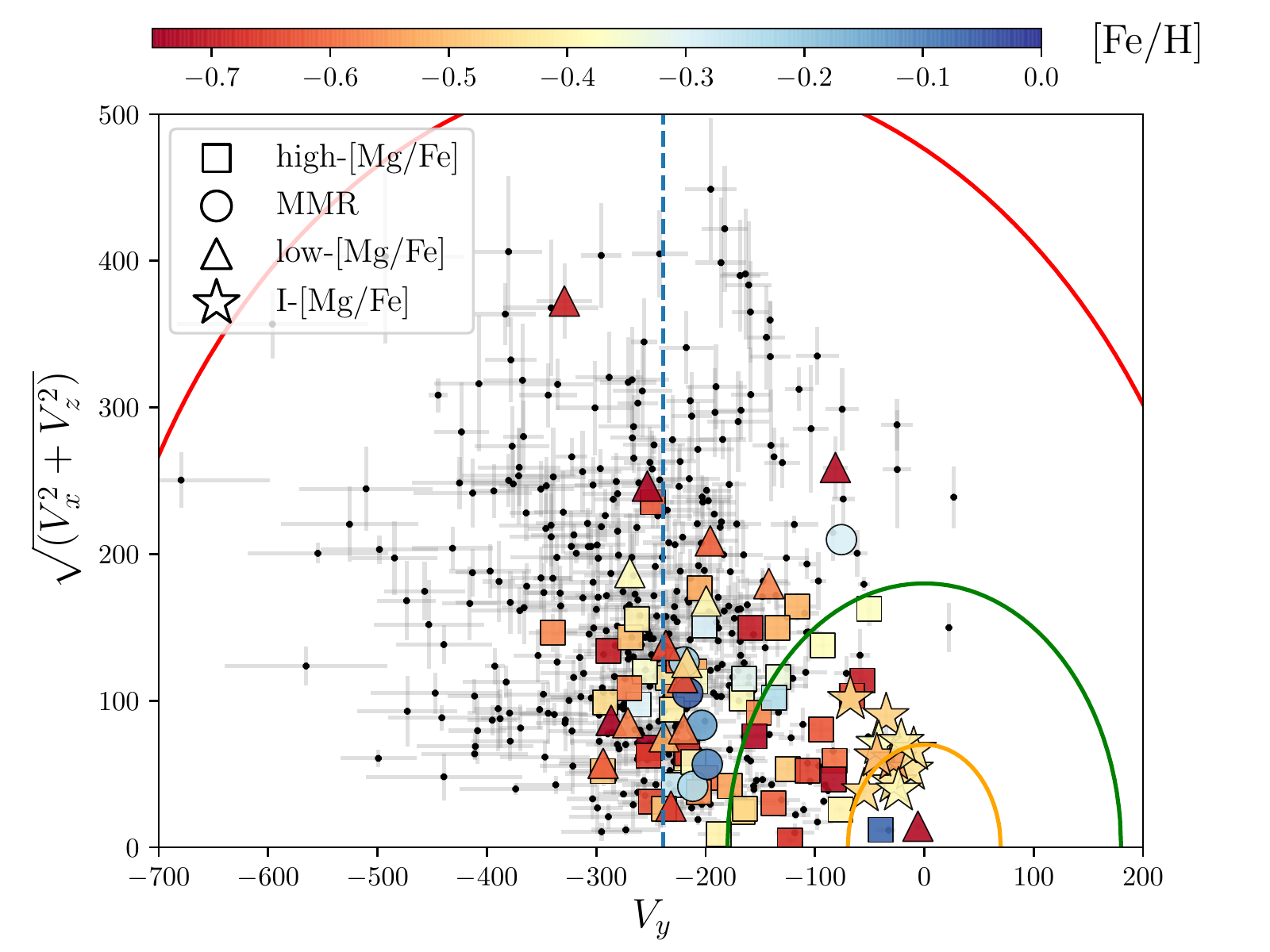}
	\caption{Toomre diagram for the metal-poor halo
sample (black dots with error bars) and metal-rich halo sample (with
colours indicating the metallicity, [Fe/H]). The metal-rich groups are highlighted with different symbols: squares (high-[Mg/Fe]), circles (MMR), triangles (low-[Mg/Fe]) and stars (I-[Mg/Fe]). Red line: Constant Galactic
rest-frame velocity of 533 km s$^{-1}$; green line: Thick-disc constant
velocity of 180 km s$^{-1}$ (Soubiran et al. 2003); orange line: Thin-disc constant velocity of 70 km $s^{-1}$
(Fuhrmann 1999). The vertical blue dashed line indicates zero Galactic rotation, V$_{lsr} = -$239 km s$^{-1}$.}
	\label{toomre}
\end{figure*}	

\begin{figure}
	\includegraphics[scale=0.6, trim= 20 0 50 0]{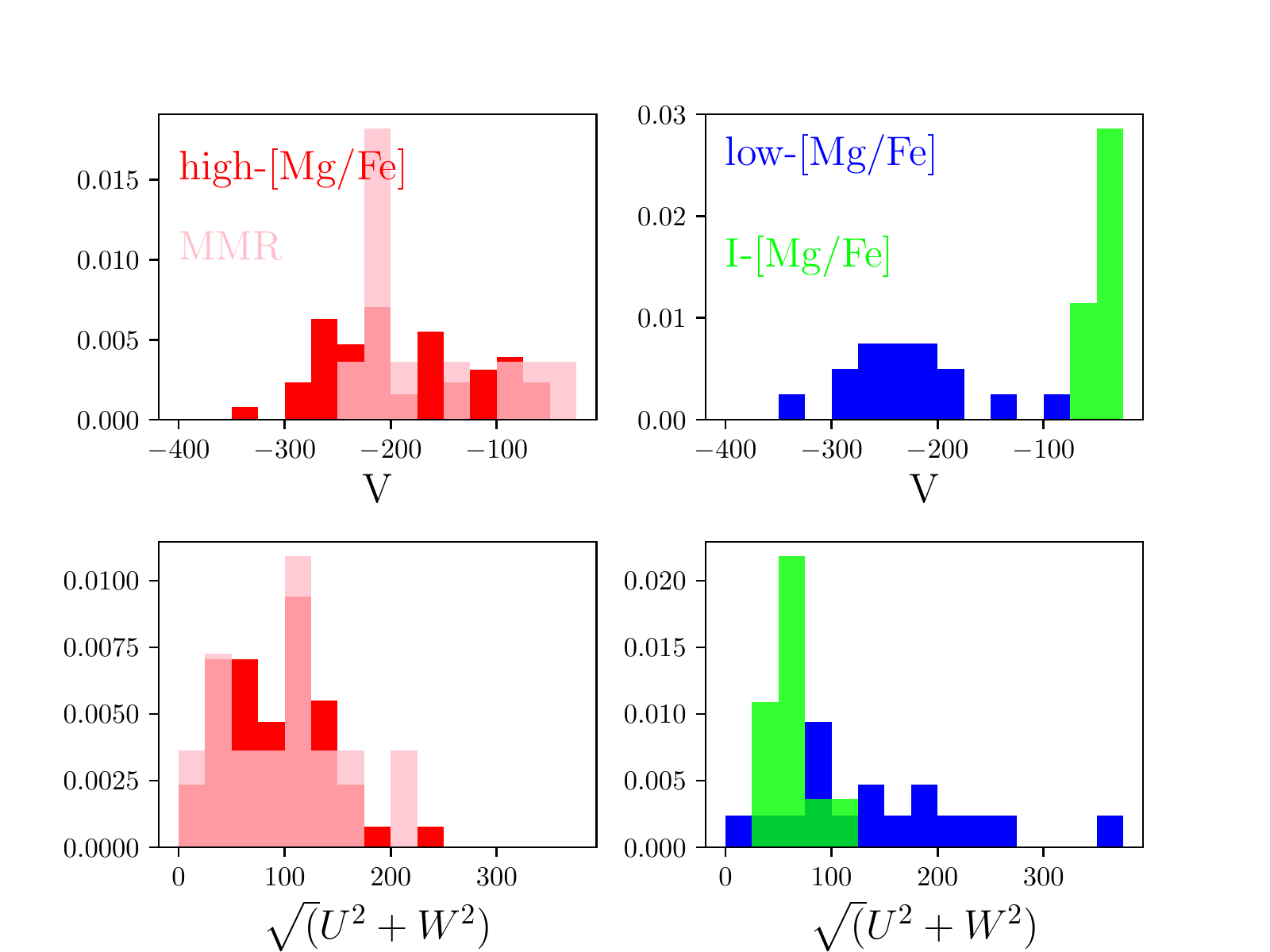}
	\caption{Top panels: Distribution
of the $V_{y}$ velocity component for stars with $-0.75 < $[Fe/H] $< -0.3$
and [Mg/Fe] $> +0.2$ (high-[Mg/Fe] + MMR -- left panel), and for stars with
$-0.75 < $[Fe/H] $< -0.3$ and [Mg/Fe] $< +0.2$ (low-[Mg/Fe] + I-[Mg/Fe] -- right
panel). Bottom panels: Distribution of the $\sqrt{V_{x}^2+V_{z}^2}$ for 
the same subsamples of stars as in the upper panels. }
	\label{toomre_histo}
\end{figure}

Figure \ref{toomre} displays the galactocentric velocity components, $V_{x}$, $V_{y}$ and $V_{z}$, derived for our sample in a Toomre diagram. The
metal-poor stars are plotted as black points, whereas the metal-rich
stellar groups are depicted using different symbols as defined in
Section 3.1, and have colours corresponding to their [Fe/H] values. The
orange and green lines indicate constant space velocities of $70 km
s^{-1}$ and $180 km s^{-1}$, respectively, which illustrate the boundaries
within which the thick-disc stars more likely dominate (Fuhrmann 1999; Soubiran et al. 2003). The exact values are not deterministic. It is only the relative ordering that contains physical information (i.e. larger asymmetric drift and velocity dispersion for the thick-disc relative to the thin-disc). Thus, they are only a reference. But we would like to consider them because these boundaries have been used in several works which aimed to characterize the thin- and thick-discs and the halo
(Venn et al. 2004, Nissen 2004, Nissen \& Schuster 2010, Bensby et al. 2014, Anders et al. 2014). The red line
shows a constant Galactic rest-frame velocity of $533 km s^{-1}$.

The entire low-[Mg/Fe] population (except for one star) and most of the high-[Mg/Fe]
and MMR stars are located outside the thick-disc boundary, i.e., they show
halo-like total space velocities, although some of the high-[Mg/Fe] and
MMR stars have lower velocities, compatible with disc kinematics.
Noticeably, the I-[Mg/Fe] stars are more concentrated in the diagram,
in contrast with the large dispersions displayed by the other stellar
groups, and overlay the velocity values delimiting the thin/thick-disc
components.

Consequently, not all our halo candidates show halo-like kinematics. However, except for the I-[Mg/Fe] group, there are high-[Mg/Fe], MMR and low-[Mg/Fe] stars moving with halo-like velocities, i.e., $V_{tot} > 200$ km/s. We check that considering only these more likely halo stars based on their kinematics we still see the [Mg/Fe] vs. [Fe/H] dichotomy up to distances $> 8$kpc .

We now examine in more statistical detail how the different metal-rich groups are distributed in
the Toomre diagram. We performed a Kolmogorov-Smirnov test to compare
the $V_{y}$ velocity component distributions for the five groups defined
above. The test carried out over the metal-rich groups results in the
strong rejection of the I-[Mg/Fe] stars belonging to the same
distribution as any of the other groups ($p$-value $< 0.001$). 

Comparing with the metal-poor sample, the statistical tests shows that
the I-[Mg/Fe] stars do not belong to the same distribution as the
metal-poor sample ($p$-value $< 0.001$). Although slightly less
significantly, the high-[Mg/Fe] stars ($p$-value = 0.02), the most
metal-rich stars ($p$-value = 0.01), and the low-[Mg/Fe] group (p-value
= 0.08) also exhibit $V_{y}$ velocity distributions that differ from that of
the metal-poor group, suggesting the latter is dominated by stars with a different kinematical behaviour than the metal-rich group.

The panels in Figure \ref{toomre_histo} are histograms of the
distribution of high-[Mg/Fe] (comparing with the MMR group of stars) and
low-[Mg/Fe] (comparing with I-[Mg/Fe] stars) for $V_{y}$ and
$\sqrt{V_{x}^{2}+V_{z}^{2}}$. Overall, this analysis reveals that the
high-[Mg/Fe] group has a large fraction of stars with prograde orbits,
but also a tail with retrograde orbits. The low-[Mg/Fe] stars are more
skewed, and distributed around $V_{y}\sim -239$ km$ s^{-1}$. The
I-[Mg/Fe] stars stand out as a group moving with a $V_{y}$ similar to the
rotation of the Galactic disc. 

The distribution of $\sqrt{V_{x}^2+V_{z}^2}$ is also different between both
groups -- high-[Mg/Fe] stars are characterized by a distribution with two
peaks at $\sim$ 50 km$ s^{-1}$ and $\sim 125$ km$ s^{-1}$. The low-[Mg/Fe] group exhibits a
broader distribution, reaching larger values, up to $\sim 300$ km$
s^{-1}$. The I-[Mg/Fe] group is characterized by low $\sqrt{V_{x}^2+V_{z}^2}$.
The median and MAD values of $V$ and $\sqrt{V_{x}^{2}+V_{z}^{2}}$ for
each group are shown in Table \ref{tab:kinematics}. 

These results highlight that the chemically distinct stellar groups are also characterized by a different kinematical signature. It is important to notice, however, that several previous works have pointed out
that a kinematical classification of stars in the different Galactic
components using the Toomre diagram do not cleanly separate them (see
Sch\"onrich \& Binney 2009, for instance). Different
chemically identified components exhibit an overlap of their velocity
distributions. In addition, comparison of the stellar velocity
components ($V_{x}$,$V_{y}$,$V_{z}$) at present time may not always be a good indicator of
the kind of orbit described by the star, and thus the Galactic
component to which it belongs, since this orbit changes with time.
Orbital parameters integrated over time are a more reliable aproach, as we
describe in the next section.

\subsection{Dynamical properties as a function of chemistry}


\begin{figure*}
	\includegraphics[scale=0.8, trim= 10 5 30 10]{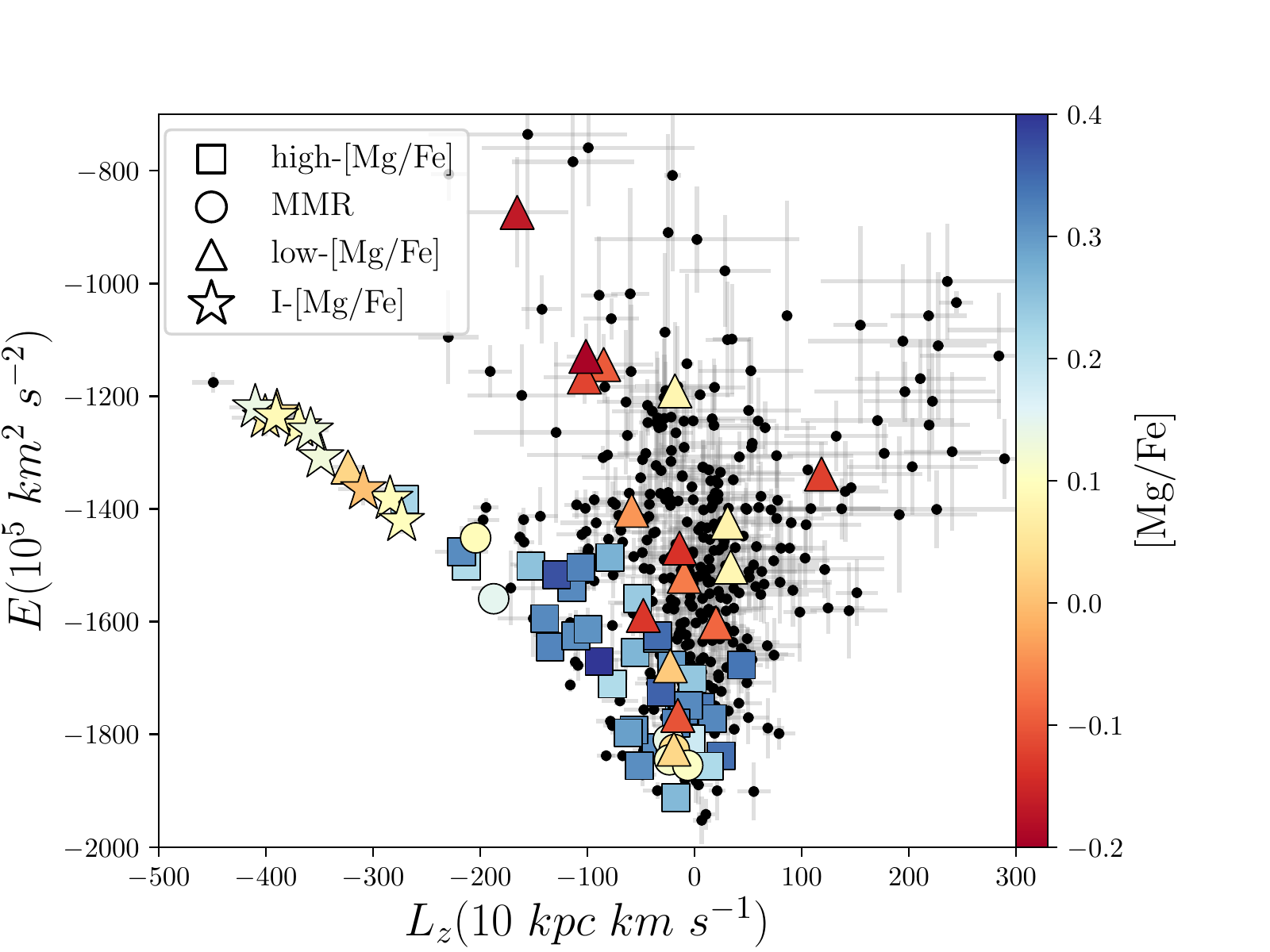}
\caption{The total orbital energy (\textit{E}) as a function of the z-component of the angular momentum ($L_z$), with colours indicating [Mg/Fe] abundance ratios for stars at [Fe/H] $> -0.75$, and black dots referring to stars with [Fe/H] $< -0.75$.}
	\label{orb_param2}
\end{figure*}

\begin{figure}
	\includegraphics[scale=0.6, trim= 20 0 0 0]{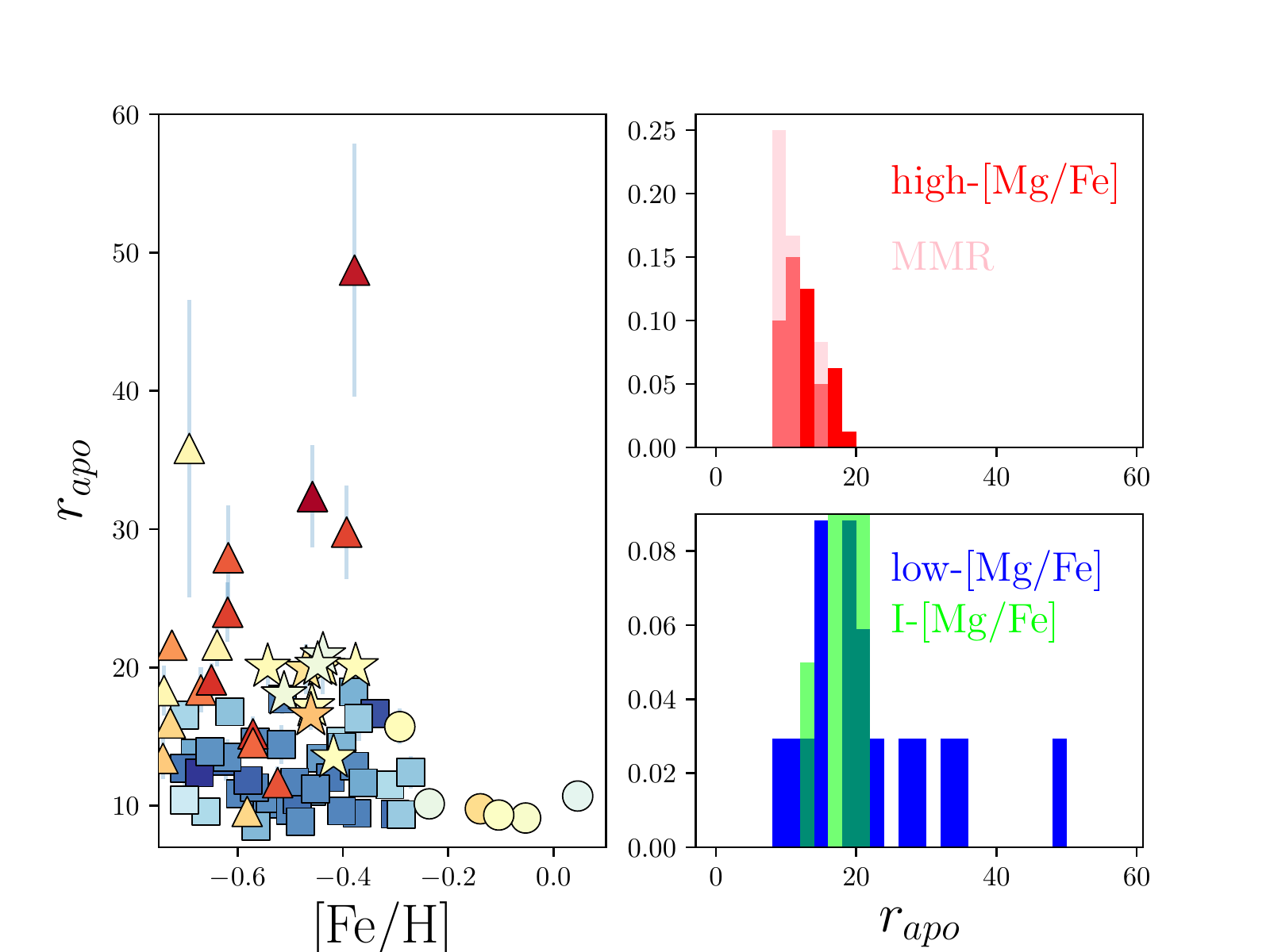}
	\includegraphics[scale=0.6, trim= 20 0 0 0]{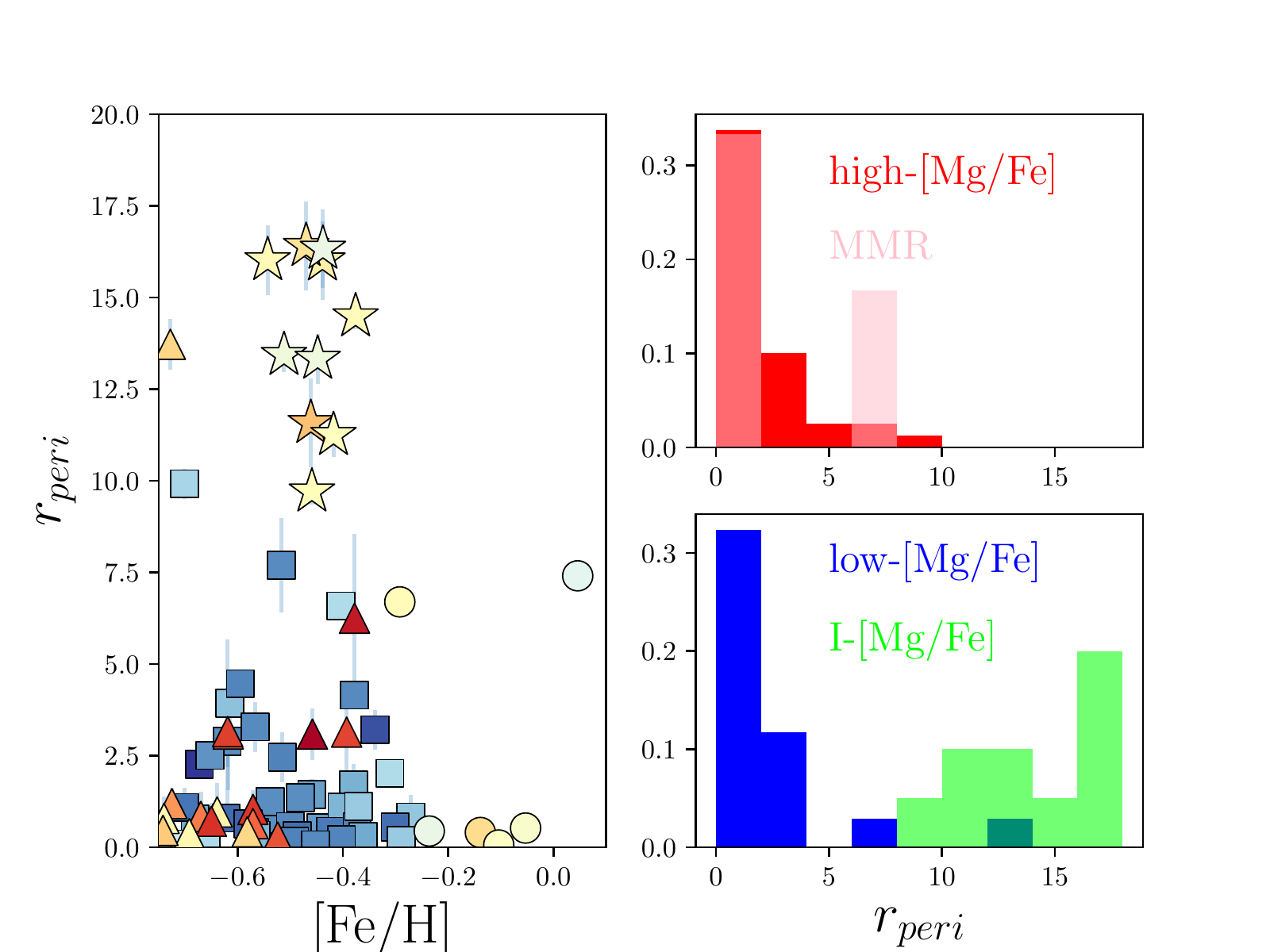}
        \caption{Left panels: Comparison of the maximum and minimum
        distance to the Galactic Centre, $r_{apo}$ and $r_{peri}$, as
        a function of [Fe/H], with colours indicating the [Mg/Fe] abundances. Right panels:
        Distribution of $r_{apo}$ and $r_{peri}$ for the high-[Mg/Fe] and
        low-[Mg/Fe] populations selected as in Figure \ref{toomre}. }
        \label{radius}
\end{figure}
	
\begin{figure}
	\includegraphics[scale=0.6, trim= 20 0 0 0]{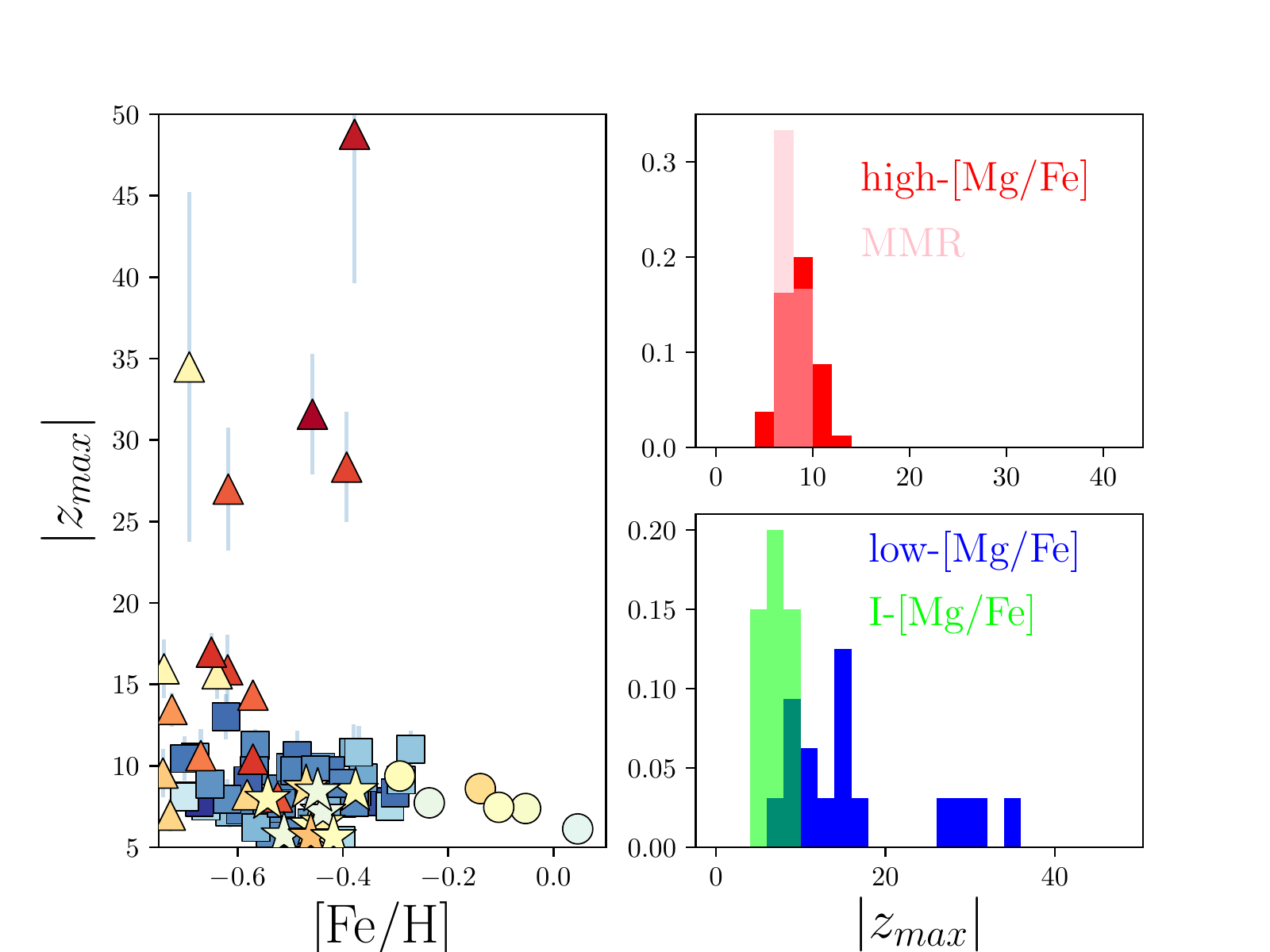}
        \caption{Left panels: Comparison of the maximum distance to the
        Galactic plane, $|z_{max}|$, as a function of [Fe/H], with colours
        indicating the
        [Mg/Fe] abundances. Right panels: Distribution of $|z_{max}|$ for
        the high-[Mg/Fe] and low-[Mg/Fe] populations selected as in Figure
        \ref{toomre}. }	 \label{zmax}
\end{figure}

\begin{figure*}
	\includegraphics[scale=0.5]{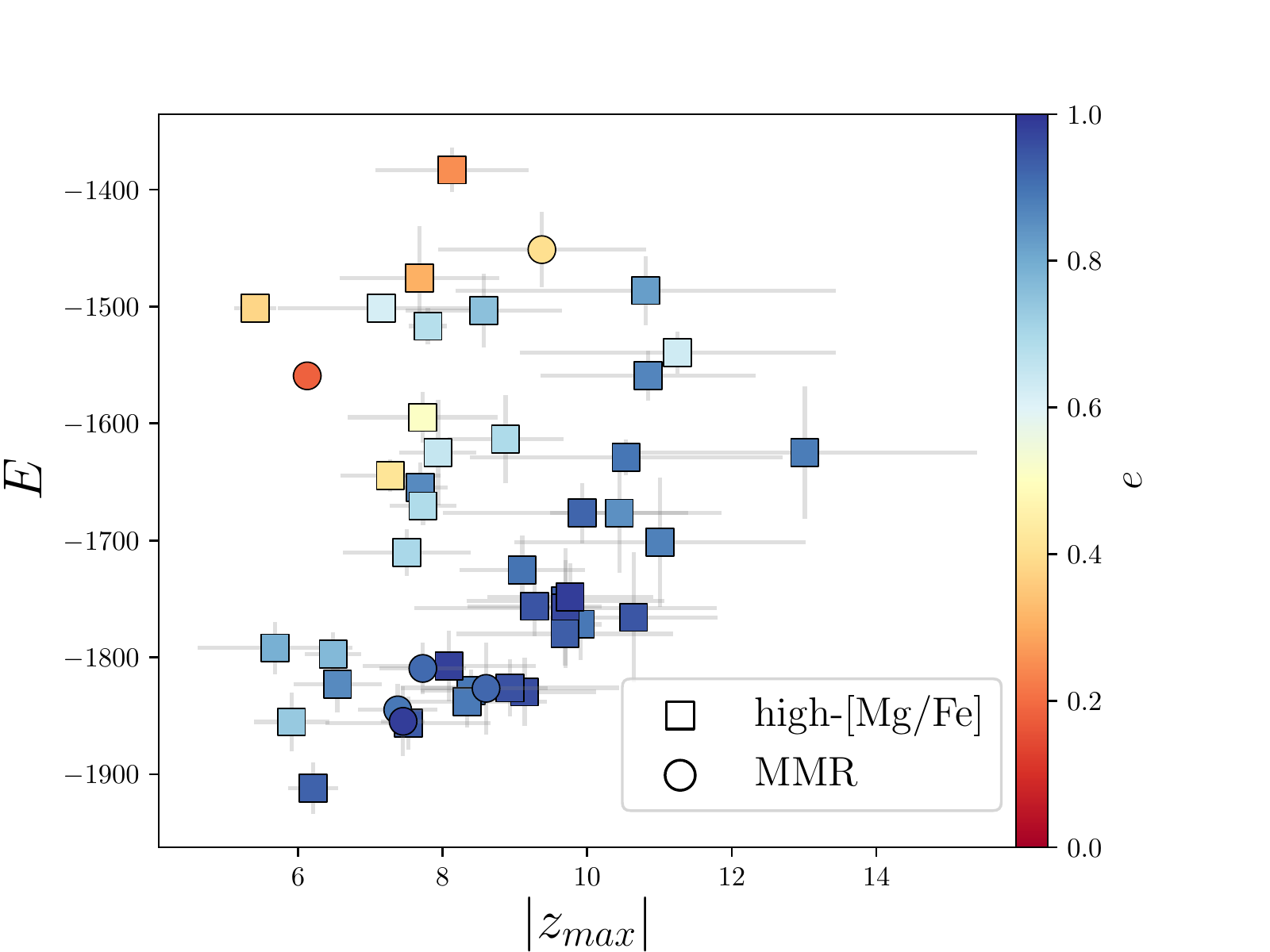}
	\includegraphics[scale=0.5]{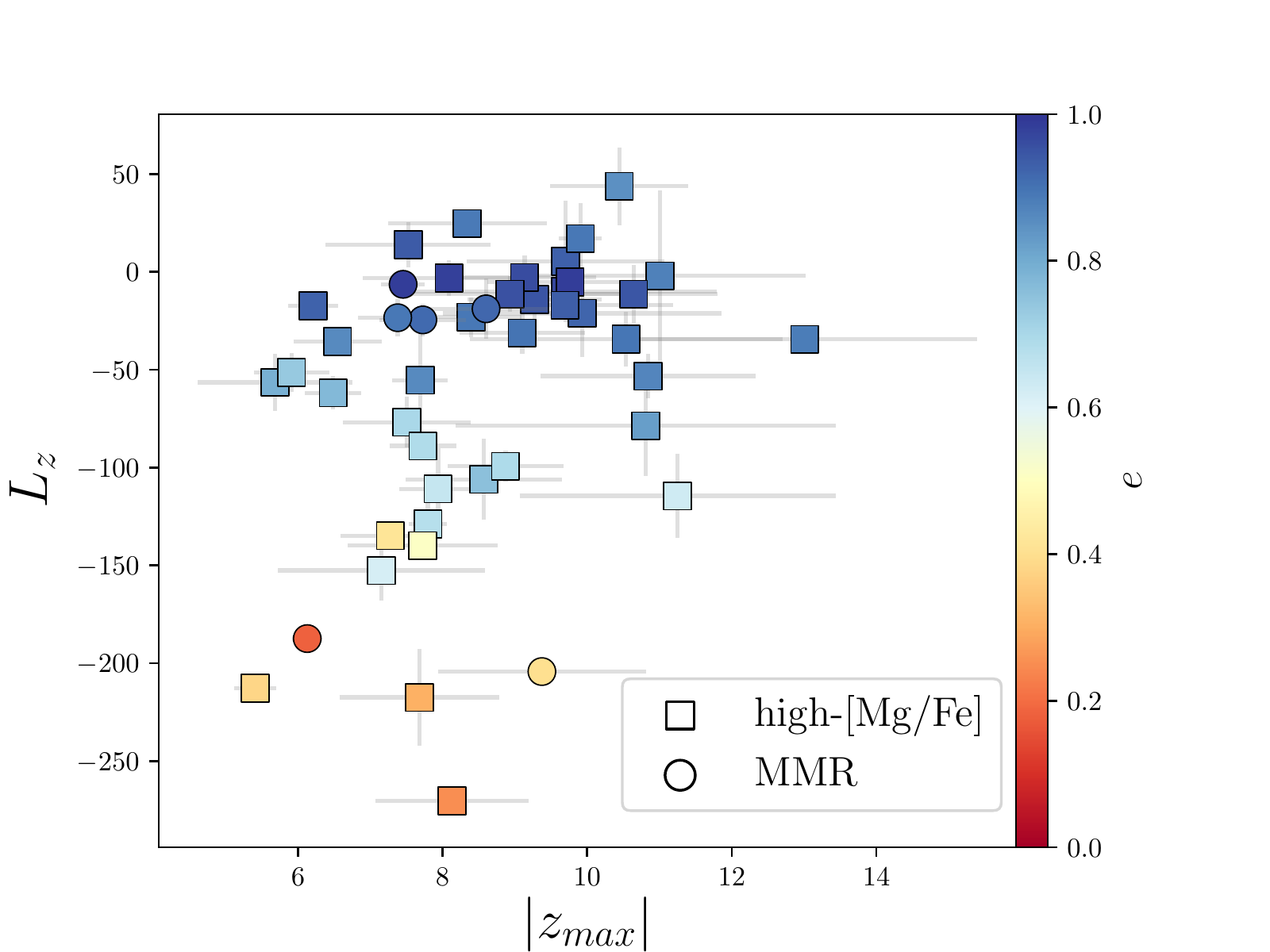}   
	\includegraphics[scale=0.5]{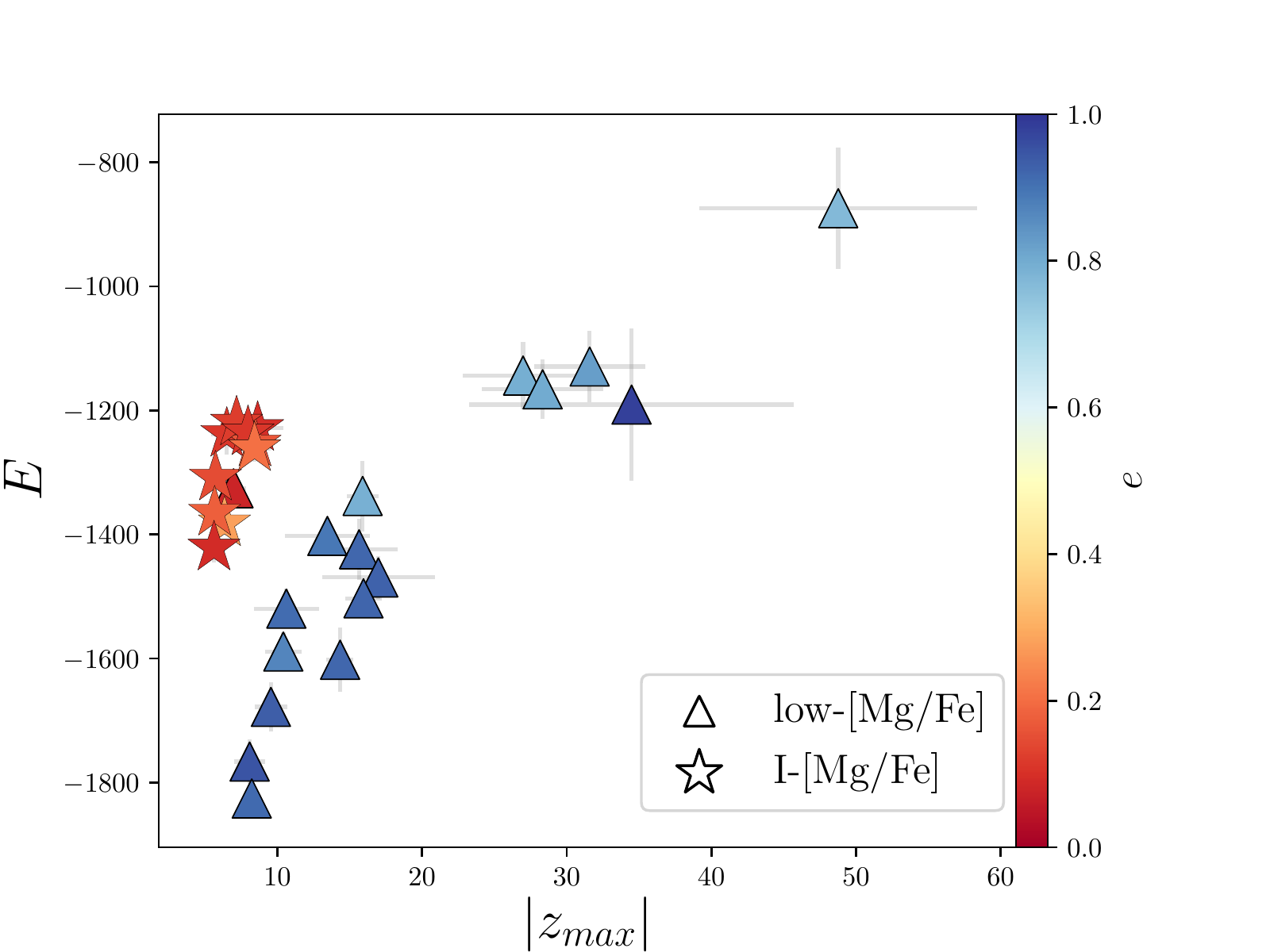}
	\includegraphics[scale=0.5]{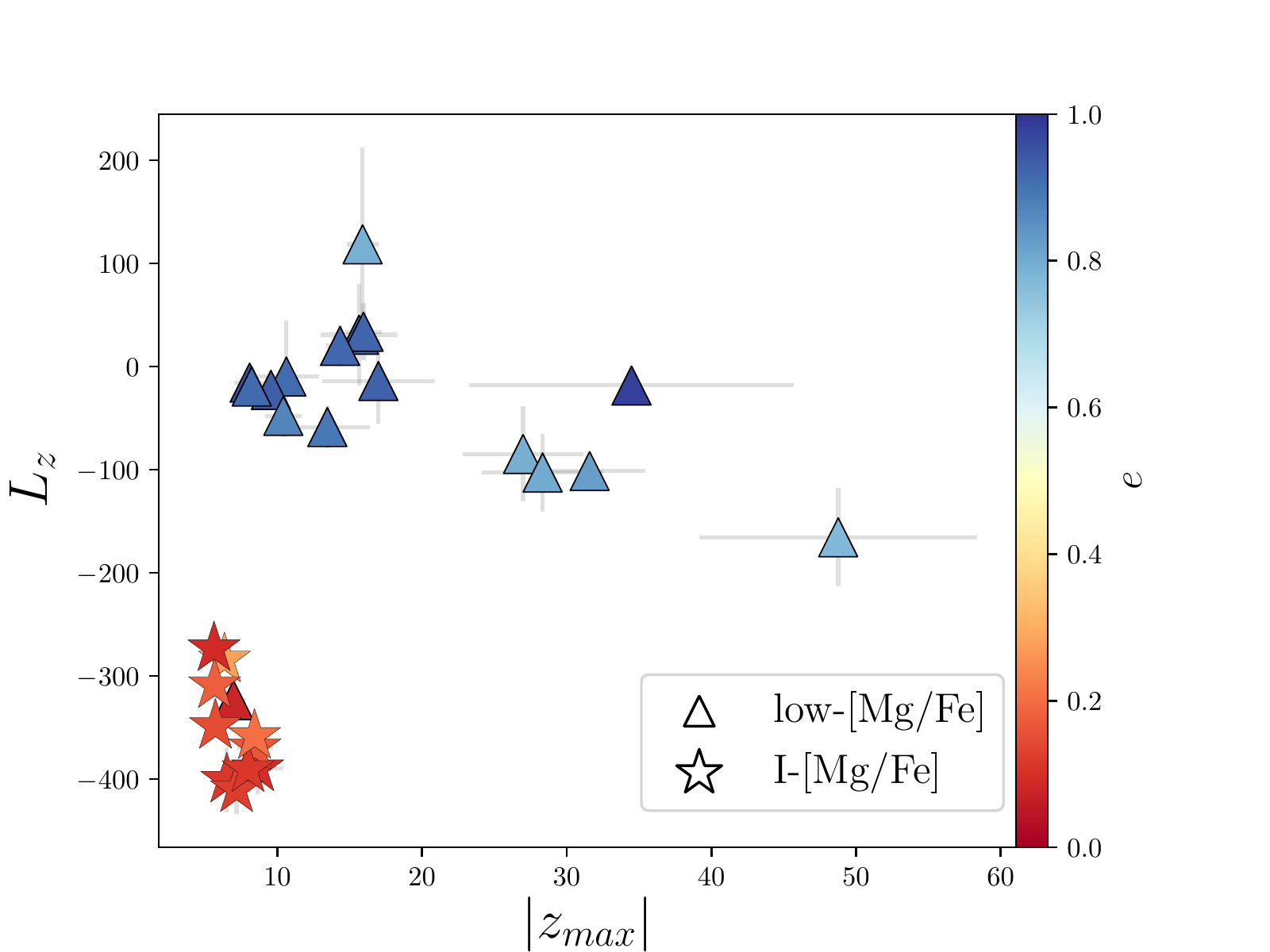}  
\caption{The total orbital energy (\textit{left panels}) and the z-component of the angular momentum (\textit{right
panels}), as a function of the maximum vertical excursion from the Galactic plane ($|z_{max}|$) for the high-[Mg/Fe] and MMR stars (\textit{top panels}, dots and triangles, respectively) and for the low-[Mg/Fe] and I-[Mg/Fe] (\textit{bottom panels}, dots and triangles, respectively), with colours indicating the orbital eccentricity.} 
\label{fig:zmax_trends}
\end{figure*}

Tables \ref{tab:dynamics1} and \ref{tab:dynamics2} shows the median values of the orbital elements derived for each group, revealing that our chemically distinct groups show also a different dynamical behaviour. To better visualize this, we plot in Figure \ref{orb_param2} our sample in action space ($E$ vs. $L_{z}$) color coded by [Mg/Fe] and with the same symbols used in previous Figures for each metal-rich stellar group, and black dots showing the metal-poor stars. Figures \ref{radius} and \ref{zmax} display the apocenters, pericenters and maximum z ($r_{apo}$, $r_{peri}$ and $|z_{max}|$) reached during the orbits as a function of [Fe/H], accompanied by histograms. Finally, Figure \ref{fig:zmax_trends} shows the trends of the total orbital energy, $E$,
and the z-component of the angular momentum, $L_z$ with $|z_{max}|$, with colours indicating the orbital eccentricity, $e$.  

The high-[Mg/Fe] stars are
characterized by lower energies, i.e., they are more bound to the Galaxy
than the low-[Mg/Fe] stars, and a more prograde median angular momentum, although some stars show no rotation or even a retrograde motion. In addition, the orbits of the high-[Mg/Fe] stars tend to remain at
lower $|z_{max}|$ than the low-[Mg/Fe] stars.

The low-[Mg/Fe] population exhibits a broader range of energies, apocenters and maximum distances from the plane, reaching higher values than the other groups, and are more symmetrically distributed around $L_z = 0$. Both the high-[Mg/Fe] and low-[Mg/Fe] populations are
characterized by large eccentricities, $e \sim 0.8$ (except a few high-[Mg/Fe] stars which follow the locus of the disc in
action space, and consequently extend toward lower eccentricities). Compared to these metal-rich stars, there are more stars within the metal-poor group moving on retrograde orbits, with lower eccentricities and higher energies.

There are trends of the total orbital energy and the z-component of the angular momentum with maximum $|z_{max}|$ reached during the orbits, for both stellar
populations. In the case of the high-[Mg/Fe] population, stars that
reach higher $|z_{max}|$ during their orbits are less bound, have an angular
momentum closer to 0. The low-[Mg/Fe]
stars exhibit the same trends for $E$, but the opposite behaviour for
$L_{z}$ with $|z_{max}|$.

It is evident that the orbits for the I-[Mg/Fe] group differ greatly from the
other groups. They are characterized by prograde orbits with low
eccenticities, reaching relatively low $|z_{max}|$ but with large energies
(i.e., they are less bound). These stars are those with the most
prograde orbits within all our stars at high $|z|$. Noticeable is that
the I-[Mg/Fe] stars reach the largest $r_{peri}$, in concordance with a
possible origin in the outer parts of the disc.

A Kolmogorov-Smirnov test shows that, comparing the four metal-rich
groups in $E$ and $|z_{max}|$, the hypothesis of belonging to the
same distribution can be rejected in the case of the high-[Mg/Fe] stars
compared with the low-[Mg/Fe] stars, but not compared with the MMR stars,
supporting the hypothesis that the MMR stars belong to the same
stellar population as the high-[Mg/Fe] stars.
This statistical test also reveals that the I-[Mg/Fe] stars have a very
different distribution (with $p$-values $< 0.001$) in every orbital
parameter compared with both the high-[Mg/Fe] and low-[Mg/Fe] populations.

In summary, our analysis reveals that the chemically distinct groups describe orbits that differ from one group to the other. Examples of stellar orbits computed for one high-[Mg/Fe], one I-[Mg/Fe] stars and one low-[Mg/Fe] stars, in the $|z|$ vs. R (radius projected onto the Galactic plane) and $y$ vs. $x$ planes, are shown in Figure \ref{orbits1} in the Appendix Section. 

\section{The origin of our high-|z| metal-rich populations.}

\label{discussion}

The metal-rich tail observed in our halo candidates sample splits into
three different [Mg/Fe] trends with [Fe/H]. This is not the first time that different sequences in the [$\alpha$/Fe] vs. [Fe/H] were detected. Nissen $\&$ Schuster (2010, 2011) identified two distinct stellar
populations in a local volume, within 350 pc centreed on the Sun. Hayes
et al. (2018a) statistically verified that these two chemically distinct
trends were the dominant populations for stars with [Fe/H] $< -0.9$.
Fern\'andez-Alvar et al. (2018) revealed the same two distinct trends,
selecting halo stars by their kinematics (large radial velocities), and
covering Galactocentric radii up to beyond 20 kpc. This is also the case
in the recent Haywood et al. (2018) work. However, this is the first
time we detect a high-/low-$\alpha$ dichotomy in metal-rich stars, [Fe/H] $> -0.75$, located at distances $> 5$ kpc from the
Galactic plane. Are we looking at the same stellar populations detected before?

\subsection{The low-[Mg/Fe] and the metal-poor stars.}

Nissen $\&$ Schuster (2010) found that their high-$\alpha$ stars tend
to have prograde orbits, whereas low-$\alpha$ stars show preferentially
retrograde orbits. Schuster et al. (2012) also found differences
between the high-$\alpha$ and low-$\alpha$ orbital parameters -- the
former showed typical values of $r_{apo} \sim$ 16 kpc, $|z_{max}| \sim$ 6-8
kpc and $e \sim$ uniformly distributed over [0.4,1.0];  the latter,
$r_{apo} \sim$ 30-40 kpc, $|z_{max}| \sim$ 18 kpc, and $e \sim$ 0.85. 

Our results are quite consistent with these values. However, our low-[Mg/Fe] sample do not exhibit preferentially retrograde orbits. Besides, their chemical abundances, locations and velocities are
consistent with the low-[Mg/Fe] population at [Fe/H] $>
-0.75$ being comprised by Sgr candidates.

Their chemical trends and their orbital parameter distributions
resemble more the ones of metal-poor stars than those displayed by the high-[Mg/Fe], MMR, or
I-[Mg/Fe] groups. Thus, our low-[Mg/Fe] group seems to be the metal-rich tip of the metal-poor
sample. However, the metal-poor group contains a larger fraction of stars on retrograde orbits.

Our metal-poor sample detected at high $|z|$
have properties more similar to the local metal-poor sample analysed by Helmi et al.
(2018), Haywood et al. (2018), and Mackereth et al. (2018), based on the APOGEE DR14 database. They claimed that most of these stars could have been
deposited by a single merger event. However, there should remain stars that
have likely been accreted from other satellite galaxies, including Sgr, which would reach larger metallicities.

\subsection{The high-[Mg/Fe] sequence: are they thick-disc stars?}

Haywood et al. (2018) obtained that stars (a subsample from APOGEE DR14
crossmatched with a high-quality Gaia data sample) sharing the red
sequence in the HRD than the high-$\alpha$,
metal-rich ([Fe/H] $> -0.8$) Nissen $\&$ Schuster population showed a
considerable number of stars with $L_z \sim$ 0 and even on retrograde
orbits. Our
analysis at high $|z|$ are in line with this. Interestingly, they also showed that chemically selected thick-disc stars
share the same red sequence and the tail to rotation decreasing to zero,
including even a few stars on retrograde orbits, pointing to a common origin
with the high-$\alpha$ metal-rich Nissen $\&$ Schuster stars. 

We found that
high-[Mg/Fe] stars exhibit more-bound orbits that move closer to the
Galactic plane than the low-[Mg/Fe] population. We also see that the $e$
distribution of the high-[Mg/Fe] group is spread towards lower
eccentricities, although the median value is $0.8\pm0.1$ (see Table
\ref{tab:dynamics1}). Ruchti et al. (2011) explored the variation of the
eccentricity distribution for stars with $|z|$ between 1 and 3 kpc
from the Galactic plane. They show that the distribution peaks at low
eccentricity, $\sim 0.4$, and the number of stars with the lowest
eccentricities diminishes, whereas the number of stars with
eccentricities larger than 0.8 increases as a function of $|z|$. The
distribution of eccentricities displayed by our high-[Mg/Fe] stars
differs from the Ruchti et al. distributions at lower $|z|$, as it peaks
at high eccentricities, and the number of stars with $e = 0.4$ is
very low, but follows the logic of the increasing fraction of highly eccentric stars trend with $|z|$ showed in their work.

The high-[Mg/Fe] chemical patterns are the same as for thick-disc
stars. However, our stars are located at larger distances from the
plane than previous works had characterized thick-disc members
(e.g., Ruchti et al. 2011; Hayden et al. 2015). Additionally, our
high-[Mg/Fe] stars exhibit a lower average angular momentum than the classically
attributed to this galactic component (e.g., Venn et al. 2004; Bensby et
al. 2014). Some of the stars show even retrograde orbits. 

However, Juric et al. (2008) showed that at $|z| \sim 5$ kpc from the 
plane the density of stars observed from Sloan data is well-reproduced
by the sum of halo and thick-disc models with comparable densities. In
addition, Sch\"onrich \& Binney (2009) predicted from their model the
existence in the Solar Neigbourhood of non-rotating thick-disc stars. If
our high-[Mg/Fe] sample have born in the same cloud that formed the
thick disc (as the similar chemistry suggests), we are detecting the
tail toward no rotation (and even counter-rotation) at high distances
from the plane.

Stars born in the disc may reach large $|z|$ in their orbits if they
were somehow perturbed into more chaotic orbits. They can also
reach large $|z|$ due to the interaction with the bar, when they
move close to the Galactic centre, as explained in Moreno et al.
(2014). However, this can only be the case for those stars showing low
angular momenta, but large eccentricities. Those stars with low
eccentricities could have not moved to such high $|z|$ in the absence of
a significant merger event. If the high-[Mg/Fe] stars in our sample
originally formed as thick-disc stars, and now have larger orbits and
lower angular momentum, the merger had to be relatively massive in order
to be able to change the stellar orbits so substantially (Qu et
al. 2011; Jean-Baptiste et al. 2017).

\subsection{The I-[Mg/Fe] stars: the discovery of a new sample of disc-heated stars.}

The I-[Mg/Fe] group is comprised by stars with [Mg/Fe] $\sim +0.1$, and
highly prograde motions with low eccentricity, reaching $\sim$ 8 kpc
from the Galactic plane, achieving relatively large $r_{apo}$ and
$r_{peri}$ during their orbits, and are significantly less bound to the
Galaxy ($E \sim -1300$). 

The positions of these stars
on the sky agree with those corresponding to the two stellar
overdensities, Triangulum/Andromeda and A13, recently pointed out by
Bergemann et al. (2018) and Hayes et al. (2018b). They found their
samples located at distances from the Galactic center $\sim 18$ kpc,
which is consistent with our derived $r_{peri}$ and $r_{apo}$ results,
$\sim$ 13 and 20 kpc. We derived a median $V_{y}$ of $-28.0\pm13.7$
km~s$^{-1}$, or
$211.0\pm13.7$ km~s$^{-1}$ if we convert it to the same reference frame as
Bergemann et al. (2018), who derived a bit lower $V$ of $195\pm25$ km~s$^{-1}$,
consistent with our result within the error bars. Inspection of the
chemical abundances showed that our sample is a little more metal rich
($<[Fe/H]> = -0.45\pm0.05$) than Bergemann et al. ($<[Fe/H]> =
-0.59\pm0.12$) and Hayes et al. ($<[Fe/H]> = -0.8$). Our [X/Fe] ratios
for O, Mg, and Ti are consistent with those derived by Bergemann et al.,
although theirs are a bit larger, likely due to the different
methodology applied. The comparison of [X/Fe] for O, Mg, Ca, K, Mn, and
Ni with those in Hayes et al. (which came from the same ASPCAP-analysed
DR14 database) shows similar values than their sample of stars at
slightly lower metallicities.

Both works have proposed that these stars were born in the disc but later heated into
their actual orbits due to tidal effects from a merger. Our derived
orbital parameters, in particular the large angular momentum, low
eccentricities, relatively low $|z_{max}|$ and large energies, are
also consistent with disc orbits heated to larger heights from the plane.
Their chemistry certainly points to formation in the Galactic outer
disc. The sample of stars considered here have not been previously
reported. Hence, our I-[Mg/Fe] group is a new sample
of stars, likely belonging to these two stellar overdensities, and the
orbital parameters derived from this analysis are compatible with the
disc origin claimed in previous works. Table \ref{tab:img} report their APOGEE\_ID as well as coordinates, radial velocities, stellar parameters and [Fe/H].

\section{Conclusions}
\label{conclusions}

We have analysed a sample of 504 halo candidate stars located at $|z| >$ 5 kpc
from the Galactic plane, selected from the APOGEE DR14 database. We
derive the MDF as a function of $|z|$, and detect a metal-rich tail up
to $|z| > 10$ kpc. We perform a chemical analysis of these metal-rich
stars using the elemental abundances derived by the ASPCAP pipeline,
and a kinematical and dynamical analysis deriving orbits from the radial
velocities also provided by the APOGEE DR14 database, spectroscopic
distances from the StarHorse code, and high-quality proper motions from Gaia DR2.

Our main conclusions are the following:

\begin{enumerate}

\item We detect metal-rich stars with $-0.75 <$ [Fe/H] $ < 0.0$ moving with halo-like kinematics ($V_{tot}$ > 200 km/s)
at distances up to $|z| > 10$ kpc from the Galactic plane. These stars show an [$\alpha$/Fe] vs. [Fe/H] dichotomy: a high-[Mg/Fe] group with a high constant
[Mg/Fe] $\sim$+0.3 and a $knee$ at [Fe/H]$\sim-0.4$; and a low-[Mg/Fe] group with
[Mg/Fe] lower than +0.1, following the decreasing [Mg/Fe] trend with
[Fe/H] that dominates the bulk of halo stars at lower metallicities. The high-[Mg/Fe] group shows a tail towards disc-like $V_{tot}$.

\item We also detect a third chemically distinct group among metal-rich stars at high-|z|, the intermediate-[Mg/Fe] stars, lying between the previous two
sequences, with [Mg/Fe] $\sim$ +0.1 and [Fe/H] $\sim -0.5$, but moving with disc-like kinematics.

\item The high-[Mg/Fe], low-[Mg/Fe], and I-[Mg/Fe] stellar groups
are also chemically distinct in other elemental species (O, Si, S, Ca,
Ti, Ni, and K). 

\item The chemical distinct stellar groups are also kinematically and dynamically different.

\item The high-[Mg/Fe] stars exhibit a distribution of $V_{y}$ with primarily slightly prograde rotation, the angular momentum decreasing to 0 as $|z_{max}|$ increases. Their
orbits are more bound to the Galaxy and closer to the Galactic plane
than the low-[Mg/Fe] stars. They typically exhibit large eccentricities,
$e = 0.8$, although the distribution extends towards lower values for a few stars. Their chemical abundances are similar to that displayed by thick-disc stars. Our results point to the high-[Mg/Fe] group as a thick-disc tail towards $L_{z} \sim 0$ at large $|z|$.

\item The low-[Mg/Fe] group has a $V_{y}$ distribution centreed around zero net rotation, 
although slightly skewed toward prograde motions. Their orbits are
less bound and reach larger $|z_{max}|$ and $r_{apo}$. Their chemical abundances, location and velocities are compatible with an origin in the Sagittarius dwarf satellite.

\item The I-[Mg/Fe] stars exhibit the largest prograde motion of the metal-rich
halo sample, and move on low-eccentricity ($e \sim$ 0.1) orbits that are 
relatively close to the Galactic plane. However, they are significantly
less bound to the Galaxy, and reach large $r_{apo}$ and $r_{peri}$
during their orbits. The I-[Mg/Fe]
stars very likely belong to the two stellar overdensities,
Triangulum-Andromeda and A13, comprising a new sample of candidates not
prevously reported.

We believe that the most likely explanation for the observed properties
of the stars in our sample is that a fraction of them
arose due to influence of a relatively massive
accretion event that had a strong interaction with a nascent thick disc, as has been suggested by a number of recent studies. Clearly, more
detailed comparisons with numerical simulations would prove
illuminating. 

\end{enumerate}

\section*{Acknowledgements}
We thank the anonymous referee for the comments provided which considerably improved this manuscript. E.F.A. would like to thank Edouard Bernard, Govind Nandakumar, \'Alvaro Rojas-Arriagada and Georges Kordopatis for very useful suggestions and discussions. E.F.A. and A.R.B. acknowledge financial support from the ANR 14-CE33-014-01. E.F.A. also thanks for partial support provided by CONACyT of M\'exico (grant 241732). J.G.F-T is supported by FONDECYTNo. 3180210. L.C. thanks for the financial supports provided by CONACyT of M\'exico (grant 241732), by PAPIIT of M\'exico (IA101217, IA101517) and by MINECO of Spain (AYA2015-65205-P). 
T.C.B. acknowledges partial support for this work from grant PHY 
14-30152; Physics Frontier Center/ JINA Center for the Evolution of the 
Elements (JINA-CEE), awarded by the US National Science Foundation. OZ and DAGH acknowledge support provided by the
Spanish Ministry of Economy and Competitiveness (MINECO) under grant
AYA-2017-88254-P. EM and APV acknowledge the grants DGAPA PAPIIT IN105916 and G100319.
APV acknowledges FAPESP for the postdoctoral fellowship 2017/15893-1.

Funding for the \texttt{GravPot16} software has been provided by the Centre national d'etudes spatiales (CNES) through grant 0101973 and UTINAM Institute of the Universit\'e de Franche-Comt\'e, supported by the R\'egion de Franche-Comt\'e and Institut des Sciences de l'Univers (INSU). Simulations have been executed on computers from the Utinam Institute of the Universit\'e de Franche- Comt\'e, supported by the R\'egion de Franche-Comt\'e and Institut des Sciences de l'Univers (INSU), and on the supercomputer facilities of the M\'esocentre de calcul de Franche-Comt\'e.

This work has made use of data from the European Space Agency (ESA)
mission {\it Gaia} (\url{https://www.cosmos.esa.int/gaia}), processed by
the {\it Gaia} Data Processing and Analysis Consortium (DPAC,
\url{https://www.cosmos.esa.int/web/gaia/dpac/consortium}). Funding for the DPAC
has been provided by national institutions, in particular the institutions
participating in the {\it Gaia} Multilateral Agreement.

Funding for the Sloan Digital Sky Survey IV has been provided by the Alfred P. Sloan Foundation, the U.S. Department of Energy Office of Science, and the Participating Institutions. SDSS acknowledges support and resources from the Center for High-Performance Computing at the University of Utah. The SDSS web site is www.sdss.org.

SDSS is managed by the Astrophysical Research Consortium for the Participating Institutions of the SDSS Collaboration including the Brazilian Participation Group, the Carnegie Institution for Science, Carnegie Mellon University, the Chilean Participation Group, the French Participation Group, Harvard-Smithsonian Center for Astrophysics, Instituto de Astrof\'isica de Canarias, The Johns Hopkins University, Kavli Institute for the Physics and Mathematics of the Universe (IPMU) / University of Tokyo, the Korean Participation Group, Lawrence Berkeley National Laboratory, Leibniz Institut f\"{u}r Astrophysik Potsdam (AIP), Max-Planck-Institut für Astronomie (MPIA Heidelberg), Max-Planck-Institut für Astrophysik (MPA Garching), Max-Planck-Institut für Extraterrestrische Physik (MPE), National Astronomical Observatories of China, New Mexico State University, New York University, University of Notre Dame, Observat\'orio Nacional / MCTI, The Ohio State University, Pennsylvania State University, Shanghai Astronomical Observatory, United Kingdom Participation Group, Universidad Nacional Aut\'onoma de M\'exico, University of Arizona, University of Colorado Boulder, University of Oxford, University of Portsmouth, University of Utah, University of Virginia, University of Washington, University of Wisconsin, Vanderbilt University, and Yale University.







\appendix

 \begin{table*}
  \caption{Stellar parameters, [Fe/H] and [Mg/Fe] abundances for our 504 halo candidates. A complete version of this table is available online.}
  \label{tab:params1}
  \begin{tabular}{lcccccccc}
  \hline
  \hline
  APOGEE ID     &       ra     &      dec     &     l     &       b      &       $T_{\rm eff} (K)$    &    $\log g$    &    [Fe/H]   &     [Mg/Fe]  \\        
  2M13381056+3823163 &  204.544 &  38.388  &  84.860   &  75.082    &  4021.1 $\pm$ 47.3  &  1.36 $\pm$  0.06   &   -0.38 $\pm$ 0.01 & 0.32   $\pm$   0.02   \\   
  2M12472244-0140359  & 191.843  & -1.677  &  300.825   &  61.178   &    4466.1  $\pm$  62.5  &  1.69 $\pm$  0.07   &   -0.72 $\pm$ 0.01 & -0.04  $\pm$  0.03 \\    
  2M14202222+5252230 &  215.093  & 52.873  &  96.269  &   59.407   &   4347.5 $\pm$  57.6  &   1.52 $\pm$  0.06   &   -0.29 $\pm$ 0.01 & 0.09  $\pm$   0.02  \\    
  2M16264146+2649592  & 246.673 &  26.833  &  45.571  &   42.432    &   4109.9  $\pm$ 50.3  &  1.10 $\pm$   0.06   &    -0.62  $\pm$ 0.01 & 0.24  $\pm$   0.03   \\   
  2M10085156+4430006  & 152.215 &  44.500  &  173.871  &   53.250   &   4075.0 $\pm$  48.4   &  1.29  $\pm$ 0.05   &   -0.06 $\pm$  0.01 &  0.17   $\pm$   0.02  \\   
  2M12161853+1452490  & 184.077  & 14.880  &  267.446  &   75.274    &   4268.9  $\pm$ 55.2  &  1.44  $\pm$ 0.06   &   -0.52 $\pm$ 0.01 & -0.10  $\pm$  0.03   \\   
... & & & & & & & &   \\ 
  \hline
  \hline
  \end{tabular}
 \end{table*}

 \begin{table*}
  \caption{Radial velocities, parallaxes, proper motions, and distances for our 504 halo candidates. A complete version of this table is available online.}
  \label{tab:params2}
  \begin{tabular}{lccccc}
  \hline
  \hline
  APOGEE ID     &     $v_{rad} (km~s^{-1})$ &   $\pi (mas)$ &  $\mu_{ra} (mas~yr^{-1})$  &  $\mu_{dec} (mas~yr^{-1})$ &  $d_{sun} (pc)$ \\        
  2M13381056+3823163 &   -90.13  $\pm$   0.08  &   0.0979  $\pm$   0.0222 &          1.125  $\pm$  0.018   &     -4.451  $\pm$  0.029    &     6157.7   $\pm$   606.8  \\   
  2M12472244-0140359  &  -66.74   $\pm$  0.17  &    0.118  $\pm$   0.026 &          3.766  $\pm$   0.054   &     -8.633  $\pm$ 0.030    &      5734.2  $\pm$  746.0 \\    
  2M14202222+5252230  &  -29.02  $\pm$   0.06  &   0.1013  $\pm$   0.0201 &          2.284  $\pm$   0.029   &     -7.050  $\pm$   0.035    &     5975.7  $\pm$  1317.2  \\    
  2M16264146+2649592  &  1.33   $\pm$   0.01  &  0.131 $\pm$   0.023 &          -2.848  $\pm$  0.034    &    -2.704  $\pm$  0.039    &     7533.3  $\pm$  663.7  \\   
  2M10085156+4430006  &  5.467  $\pm$    0.002  &  0.227  $\pm$   0.049 &          -1.715  $\pm$  0.055    &    -2.734  $\pm$  0.071   &      7536.5   $\pm$   690.6  \\   
  2M12161853+1452490  &  111.80   $\pm$   0.20   &   0.0934  $\pm$   0.0507  &         -5.598  $\pm$  0.0900     &    -5.704  $\pm$  0.086    &     6486.6  $\pm$   618.7  \\   
... & & & & &  \\ 
  \hline
  \hline
  \end{tabular}
 \end{table*}

\begin{table*}
\caption{I-[Mg/Fe] stars identified as belonging to the Triangulum/Andromeda and A13 stellar overdensities. }
\label{tab:img}
\begin{tabular}{lcccccc}
\hline
\hline
 APOGEE ID     &     ra    &     dec    &  $v_{rad} (km~s^{-1})$   &    $T_{\rm eff} (K)$    &    $\log g$    &    [Fe/H]    \\
  2M03255238+0921586 & 51.468  &  9.366  &  27.46  $\pm$  0.17 & 4532.8 $\pm$ 76.5 & 2.07 $\pm$   0.07 & $-$0.46 $\pm$ 0.01 \\
  2M07421181+6645232 & 115.549 &  66.756 &  $-$49.98  $\pm$ 0.11 & 4412.3 $\pm$ 78.8 & 1.60 $\pm$   0.08   & $-$0.44 $\pm$ 0.01 \\
  2M07510819+5232100 & 117.784 &  52.536 &  $-$24.26 $\pm$  0.00 & 4238.6 $\pm$ 63.6 & 1.32 $\pm$   0.07  & $-$0.47 $\pm$ 0.01 \\
  2M08000496+5223185 & 120.0207 & 52.388 &  8.49   $\pm$  0.00 & 4071.7 $\pm$ 55.3 & 1.09  $\pm$   0.06  & $-$0.38 $\pm$ 0.01 \\
  2M08031715+4305211 & 120.821 &  43.089 &  $-$38.92 $\pm$  0.11 & 4259.6 $\pm$ 61.3 & 1.37  $\pm$  0.07  & $-$0.44 $\pm$ 0.01 \\
  2M08170874+3252000 & 124.286 &  32.867 &   54.83  $\pm$  0.21 & 4592.5 $\pm$ 76.8 & 2.01  $\pm$  0.07 & $-$0.54  $\pm$ 0.01 \\
  2M08455950+1141225 & 131.498 &  11.690 &  69.44  $\pm$  0.07 & 4089.4 $\pm$ 49.5 & 1.11 $\pm$   0.06  & $-$0.51 $\pm$ 0.01 \\
  2M09083889+4130457 & 137.162 &  41.513 &  $-$59.24 $\pm$  0.08 & 4275.6 $\pm$ 65.1 & 1.45 $\pm$   0.07  & $-$0.45 $\pm$ 0.01 \\
  2M09233417-0657105 & 140.892 &  $-$6.953 &  115.94 $\pm$  0.11 & 4290.2 $\pm$ 62.3 & 1.44 $\pm$   0.07 & $-$0.46 $\pm$ 0.01 \\
  2M10144610+0233578 & 153.692 &  2.566  &  53.07  $\pm$  0.00 & 4556.5 $\pm$ 78.4 & 1.85 $\pm$   0.07  & $-$0.42  $\pm$ 0.01 \\
  \hline
  \hline
  \end{tabular}
  \end{table*}

 \begin{table*}
  \caption{Median and median absolute deviation (MAD) of the
Galactocentric velocity components for the
metal-rich halo stars, separated into four groups, as explained in the
text --  high-[Mg/Fe], most metal-rich (MMR), low-[Mg/Fe], and intermediate-Mg (I-[Mg/Fe]).}
  \label{tab:kinematics}
  \begin{tabular}{lcccccc}
  \hline
  \hline
  &  & $V_{y} (km~s^{-1})$  &  &  & $\sqrt{V_{x}^{2}+V_{z}^{2}} (km~s^{-1})$ &   \\
  \hline
  \hline
  \textbf & $all$ &  $d_{BJ}<40\%$ & $\pi_e<50\%$  & $all$ & $d_{BJ}<40\%$ & $\pi_e<50\%$  \\
  \hline
  \hline
   \textbf{high-[Mg/Fe]} & & & & & & \\
  \hline
  median & $-$207.8 & $-$209.7 & $-$171.9 & 95.6 & 100.1 & 86.2         \\
  MAD & 50.1 & 48.1 & 51.5 & 38.6 & 40.8 & 33.6  \\
  \hline
  \hline
   \textbf{MMR} & & & & & &  \\
  \hline
  median & $-$201.1 & $-$199.8 & $-$201.1 & 102.1 & 103.5 & 102.1  \\
  MAD & 18.4 & 14.2 & 27.8 & 57.4 & 33.6 & 49.0 \\
  \hline
  \hline
   \textbf{low-[Mg/Fe]} & & & & & & \\
  \hline
  median & $-$239.6 & $-$228.9 & $-$242.4 & 115.8 & 133.2 & 101.2  \\
  MAD & 36.0 & 72.1 & 27.7 & 58.6 & 85.4 & 58.6 \\
  \hline
  \hline
   \textbf{I-[Mg/Fe]} & & & & & &  \\
  \hline
  median & $-$28.0 & $-$16.8 & $-$33.5 & 61.9 & 59.6 & 45.8  \\
  MAD & 13.7 & 5.7 & 15.4 & 9.9 & 9.5 & 7.3  \\

  \hline
  \hline
  \end{tabular}
 \end{table*}

 \begin{table*}
  \caption{Median and median absolute deviation (MAD) of the orbital parameters $E$, $L_z$, and $e$ for the
metal-rich halo stars, separated into four groups, as explained in the
text --  high-[Mg/Fe], most metal-rich (MMR), low-[Mg/Fe], and intermediate-[Mg/Fe] (I-[Mg/Fe]).}
  \label{tab:dynamics1}
  \begin{tabular}{lccccccccc}
  \hline
  \hline
    &  & $E (10^5~km^{2}~s^{-2})$ &  &  & $L_z (10~kpc~km~s^{-1})$ &  &  & $e $ &   \\
  \hline
  \hline
  \textbf & $all$ &  $d_{BJ}<40\%$ & $\pi_e<50\%$  & $all$ & $d_{BJ}<40\%$ & $\pi_e<50\%$ & $all$ &  $ d_{BJ}<40\%$ & $ \pi_e<50\%$  \\
  \hline
  \hline
  \textbf{high-[Mg/Fe]} & & & & & & & & & \\
  \hline
  median  & $-$1736.8 & $-$1752.3 & $-$1781.9 & $-$43.5 & $-$35.1 & $-$52.3 & 0.8 & 0.8 & 0.8\\
  MAD & 91.9 & 76.1 & 82.4 & 40.1 & 28.8 & 43.4 & 0.1 & 0.1 & 0.1 \\
  \hline
  \hline
  \textbf{MMR} & & & & & & & & &  \\
  \hline
  median & $-$1809.3 & $-$1818.0 & $-$1790.6 & $-$23.4 & $-$24.0 & $-$18.9 & 0.9 & 0.9 & 0.9 \\
  MAD & 45.2 & 27.1 & 89.1 & 21.4 & 19.5 & 30.0 & 0.1 & 0.1 & 0.1  \\
  \hline
  \hline
   \textbf{low-[Mg/Fe]} & & & & & & & & &  \\
  \hline
  median  & $-$1469.3 & $-$1545.9 & $-$1511.3 & $-$19.3 & $-$21.0 & $-$14.4 & 0.9 & 0.9 & 0.9 \\
  MAD & 132.7 & 181.9 & 99.8 & 39.3 & 16.1 & 39.3 & 0.03 & 0.03 & 0.01  \\
  \hline
  \hline
   \textbf{I-[Mg/Fe]} & & & & & & & & &  \\
  \hline
  median  & $-$1261.2 & $-$1249.6 & $-$1334.7 & $-$358.3 & $-$379.5 & $-$323.6 & 0.1 & 0.1 & 0.2  \\
  MAD & 41.3 & 20.7 & 36.8 & 42.4 & 25.8 & 32.2 & 0.03 & 0.02 & 0.1  \\

  \hline
  \hline
  \end{tabular}
 \end{table*}

 \begin{table*}
  \caption{Median and median absolute deviation (MAD) of the orbital parameters $|z_{max}|$, $r_{apo}$, and $r_{peri}$ for the
metal-rich halo stars, separated into four groups, as explained in the
text --  high-[Mg/Fe], most metal-rich (MMR), low-[Mg/Fe], and intermediate-[Mg/Fe] (I-[Mg/Fe]).}
  \label{tab:dynamics2}
  \begin{tabular}{lccccccccc}
  \hline
  \hline
    &   & $|z_{max}| (kpc)$ &  &  & $r_{apo} (kpc)$  &   &  & $r_{peri} (kpc)$  &      \\
  \hline
  \hline
  \textbf & $all$ &  $d_{BJ}<40\%$ & $\pi_e<50\%$  & $all$ & $d_{BJ}<40\%$ & $\pi_e<50\%$ & $all$ &  $ d_{BJ}<40\%$ & $ \pi_e<50\%$  \\
  \hline
  \hline
  \textbf{high-[Mg/Fe]} & &   &   &  &  &  &  &   &   \\
  \hline
  median  &  7.8 & 8.5 & 7.5  & 11.4 & 11.7 & 10.7 & 0.94 & 0.94 & 1.2 \\
  MAD &  1.5 & 1.3 & 1.2  & 2.0 & 1.9 & 1.7 & 0.70 & 0.68 & 0.92  \\
  \hline
  \hline
  \textbf{MMR} & & & & & & & & & \\
  \hline
  median &  7.4 & 7.6 & 8.6  & 9.8 & 9.6 & 9.8 & 0.52 & 0.48 & 0.82 \\
  MAD &  1.7 & 1.2 & 2.4  & 0.91  & 0.51 & 1.1 & 0.34 & 0.36 & 0.64  \\
  \hline
  \hline
   \textbf{low-[Mg/Fe]} & & & & & & & & & \\
  \hline
  median  &  14.3 & 10.0 & 13.9  & 18.4 & 15.6 & 18.3 & 0.96 & 0.62 & 0.90 \\
  MAD &  3.9 & 2.5 & 2.7  & 3.8 & 3.3 & 3.5 & 0.50 & 0.30 & 0.27 \\
  \hline
  \hline
   \textbf{I-[Mg/Fe]} & & & & & & & & &  \\
  \hline
  median  &  6.5 & 6.8 & 6.0  & 19.9 & 20.1 & 17.9 & 13.4 & 14.7 & 12.0 \\
  MAD &  0.9 & 0.78 & 0.73  & 0.98 & 0.38 & 0.43 & 2.6 & 1.5 & 1.9 \\

  \hline
  \hline
  \end{tabular}
 \end{table*}

\begin{figure*}
	\includegraphics[scale=0.5, trim= 0 0 0 0]{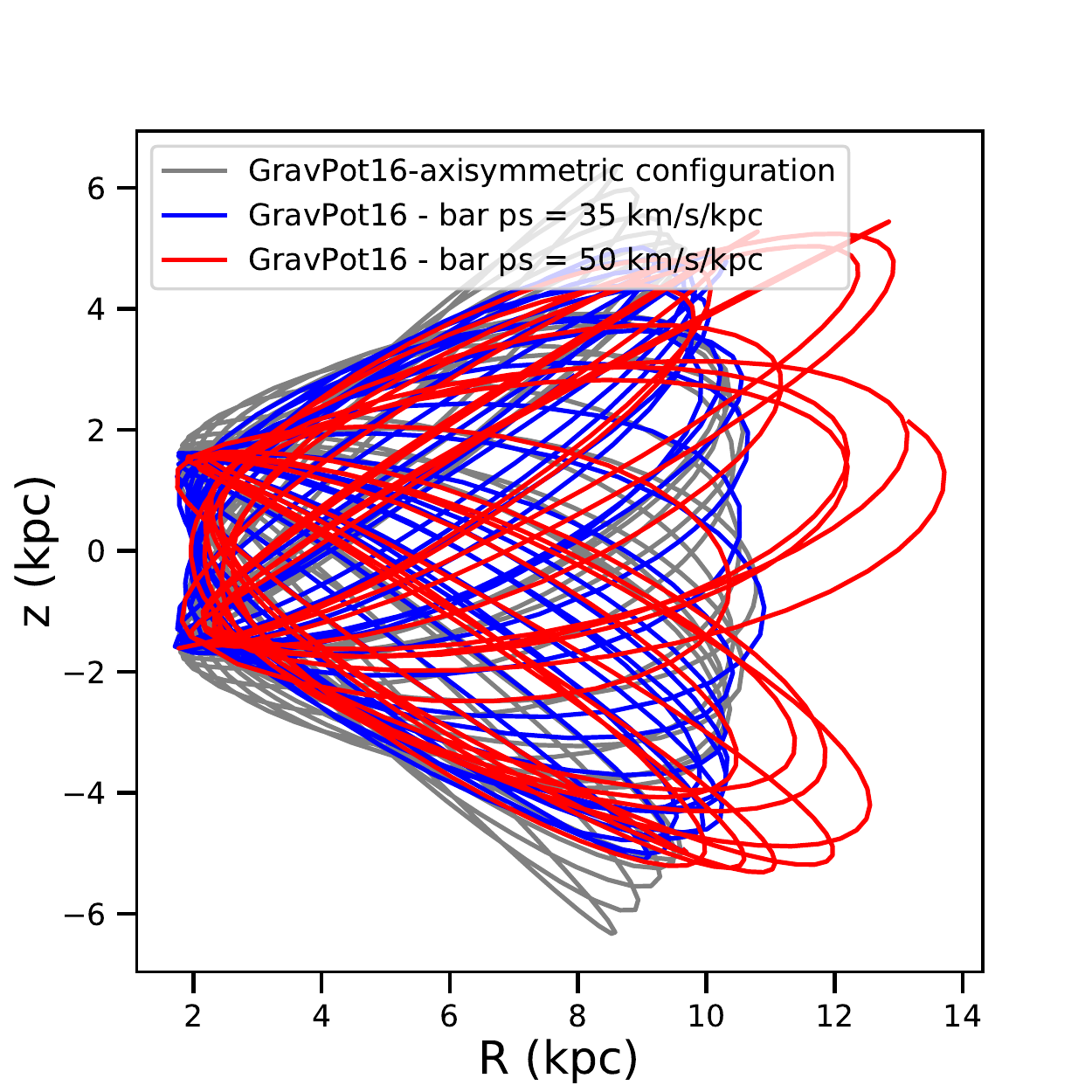}   
	\includegraphics[scale=0.5, trim= 0 0 0 0]{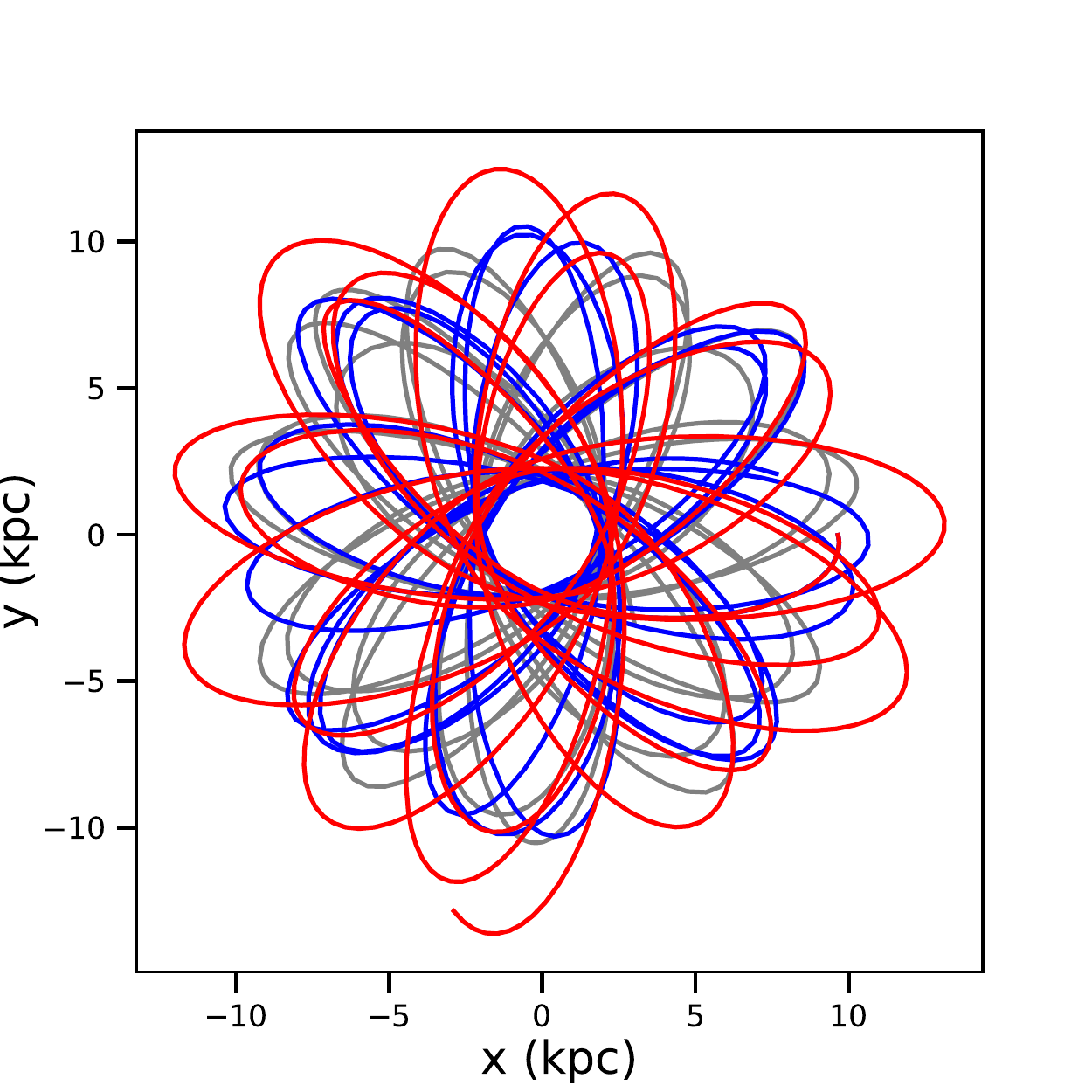}   
	\includegraphics[scale=0.5, trim= 0 0 0 0]{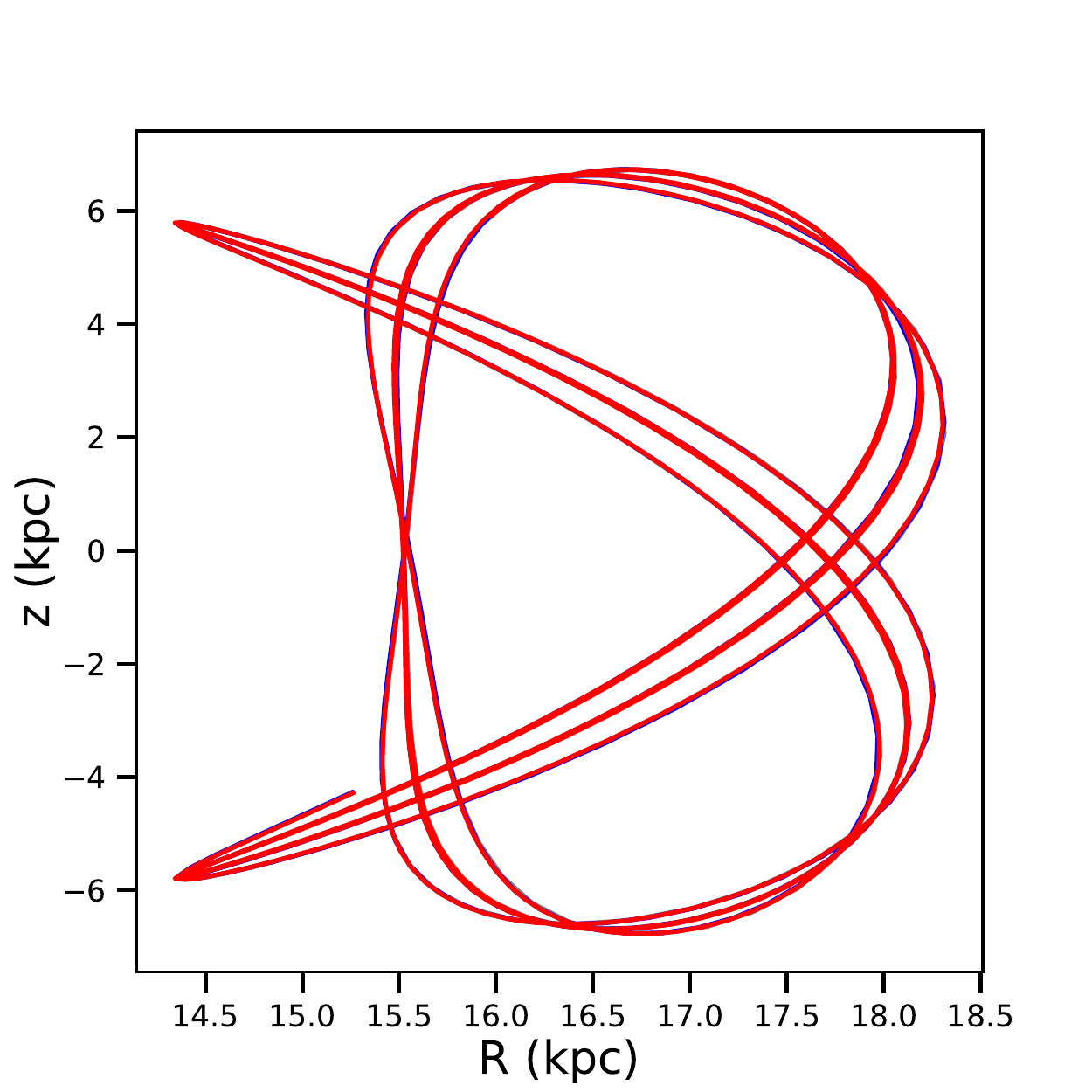}   
	\includegraphics[scale=0.5, trim= 0 0 0 0]{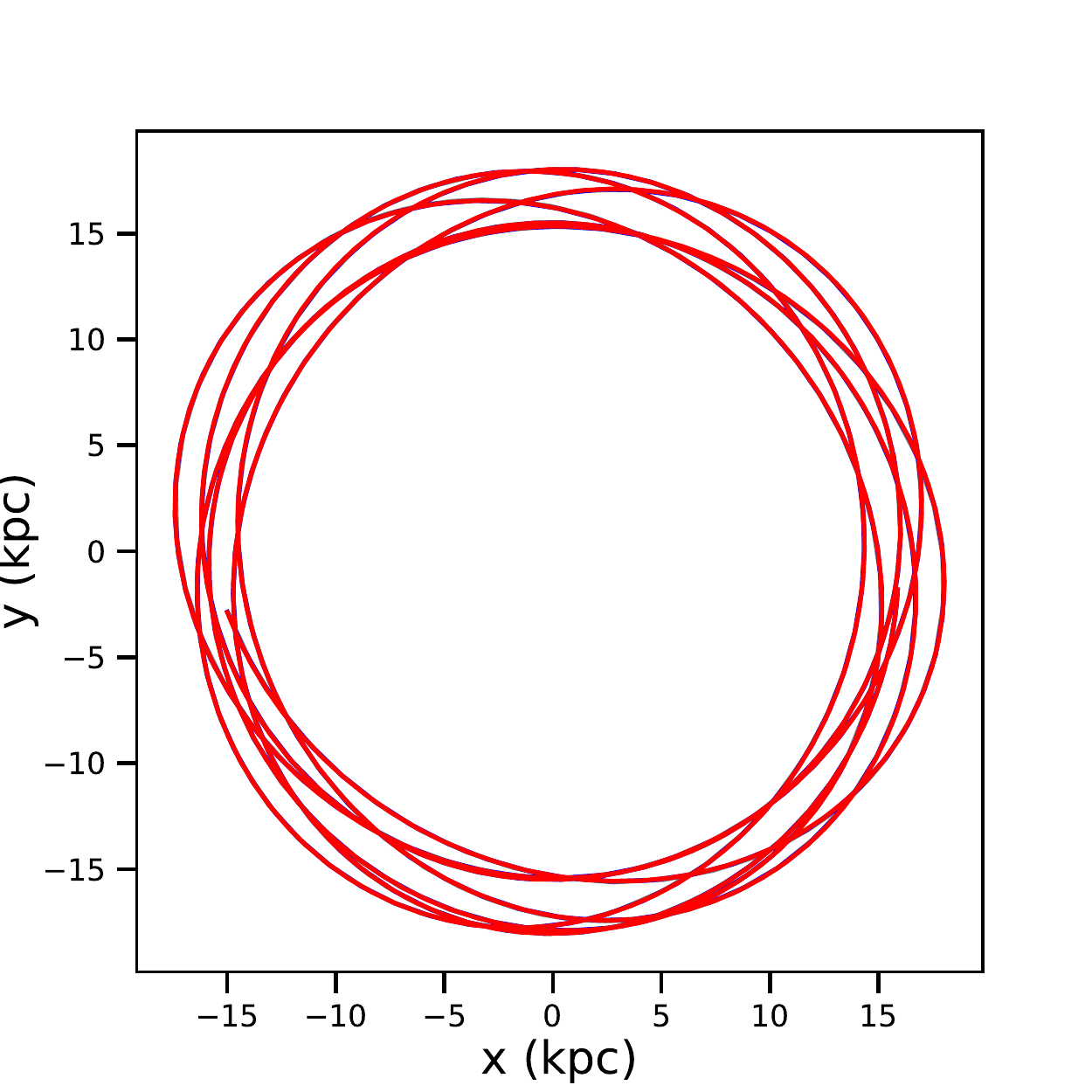}   

	\includegraphics[scale=0.5, trim= 0 0 0 0]{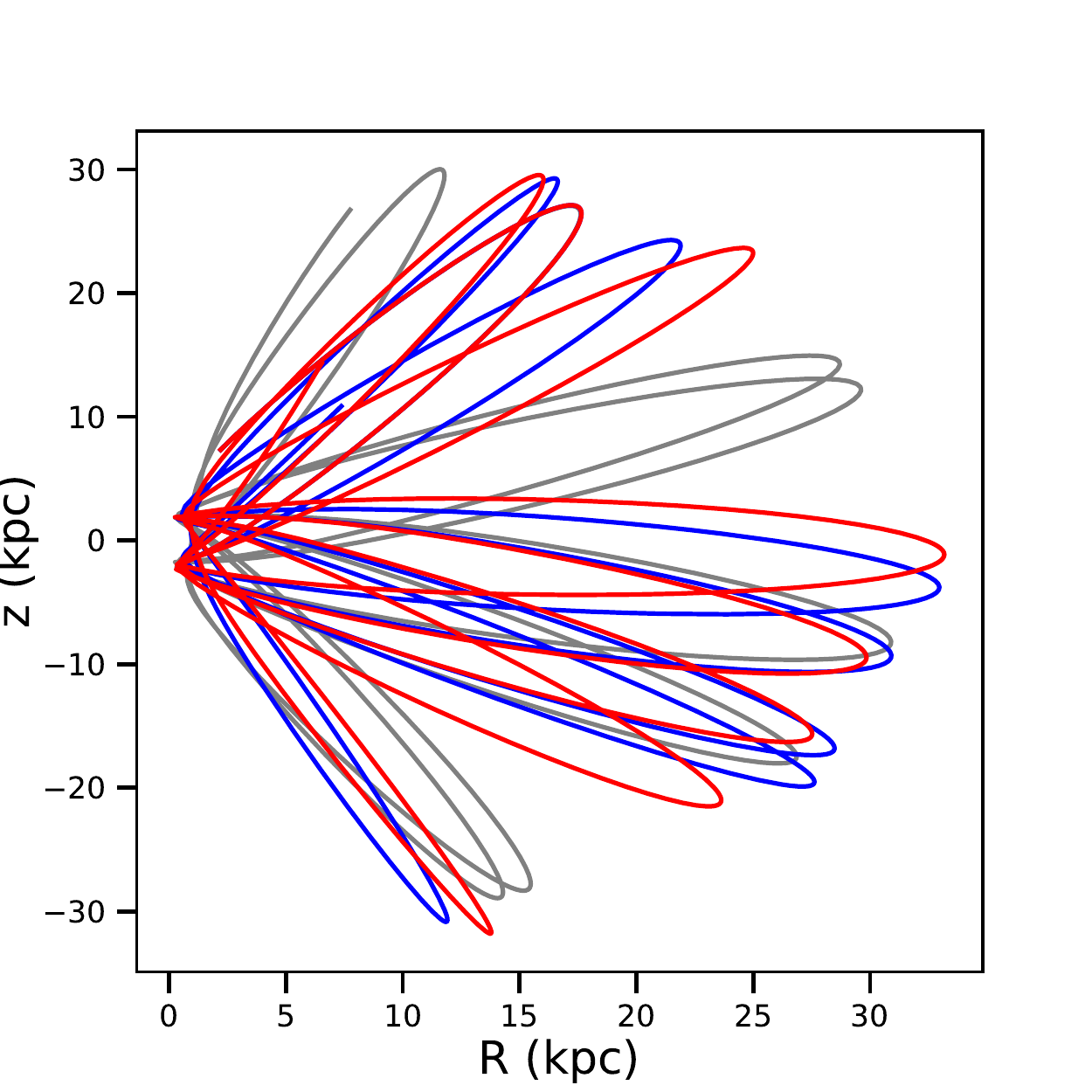}   
	\includegraphics[scale=0.5, trim= 0 0 0 0]{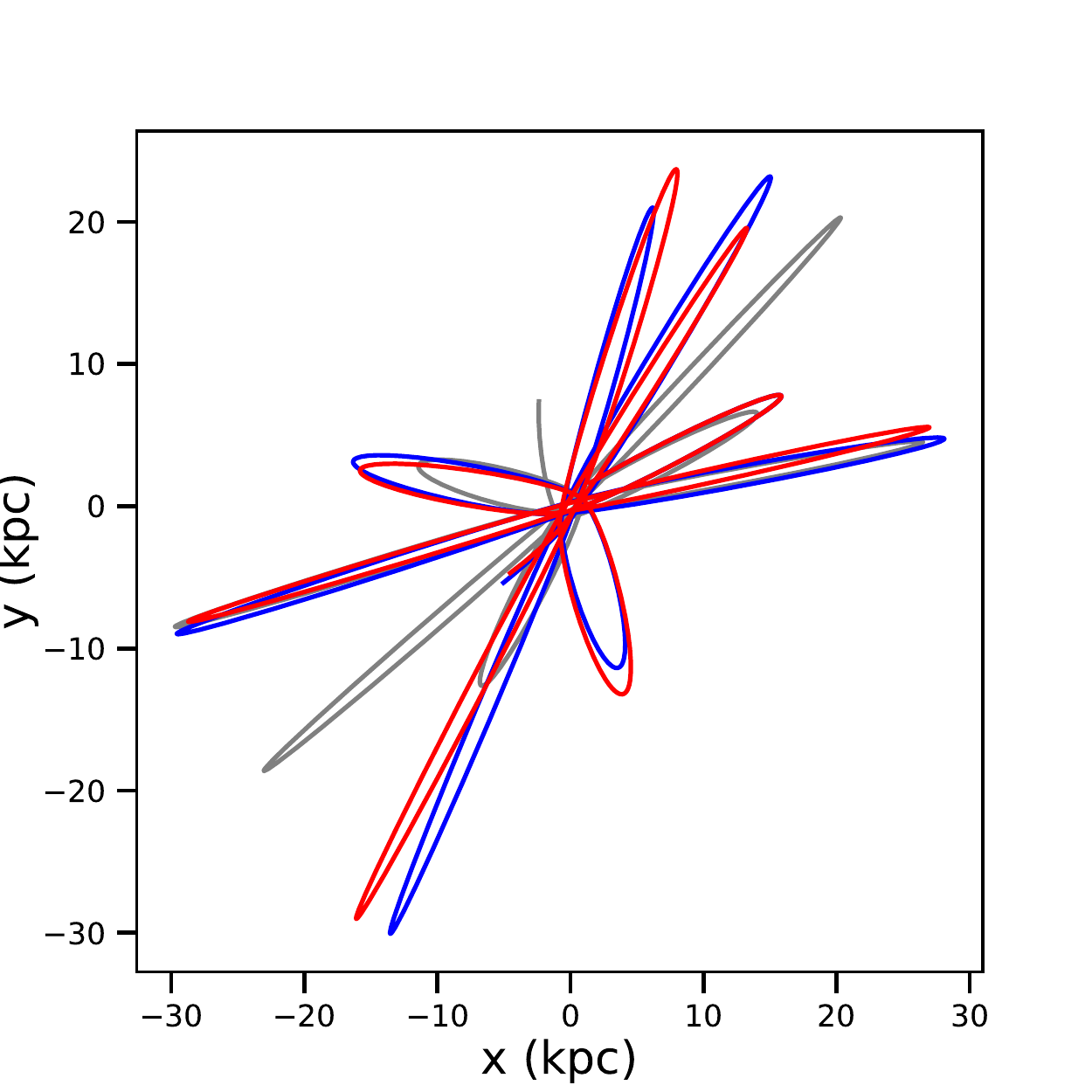}   

	\caption{Computed orbits using an axisymmetric (grey) and non-axisymmetric configuration of \texttt{GravPot16} potential model at different angular velocities for the bar (bar ps = $35 km~s^{-1}~kpc^{-1}$ -- red, and $50 km~s^{-1}~kpc^{-1}$ -- blue), for stars in our halo sample with
[Fe/H] $> -0.75$. First row: A high-[Mg/Fe] star, <$E>$ = $-1449.064x10^{5}~km^{2}~s^{-2}$, $<L_z> = 70.304x10 kpc~km~s^{-1}$, $|z_{max}|$ = 5.546 kpc, and $<ecc>$ = 0.79. 
Second row: An I-[Mg/Fe] star, $<E>$ > $-1300x10^{5}~km^{2}~s^{-2}$, $<L_z>$ < $-200x10 kpc~km~s^{-1}$, $<ecc>$ <
0.4. Third row: A star with $E_{m} >
-1000x10^{5}~km^{2}~s^{-2}$ and $<L_z> > 200x10 kpc~km~s^{-1}$. }
	\label{orbits1}
\end{figure*}


\bsp	
\label{lastpage}
\end{document}